\documentclass[letterpaper,twocolumn,10pt]{article}
\usepackage{usenix2019_v3}

\usepackage{tikz}
\usepackage{amsmath}
\usepackage[english]{babel}
\usepackage{blindtext}
\usepackage{mathrsfs}
\usepackage{amsfonts}
\usepackage{subfigure}
\usepackage{amsthm}
\usepackage{multirow}
\usepackage{booktabs}
\usepackage{enumitem}
\usepackage{mathtools}
\usepackage{hyperref}
\usepackage{tabularx}
\usepackage{listings}
\definecolor{oiReddishPurple}{HTML}{CC79A7}
\definecolor{oiBluishGreen}{HTML}{009E73}
\definecolor{oiSkyBlue}{HTML}{56B4E9}

\usepackage{tikz}
\usepackage{bbding}
\usepackage[ruled,vlined,linesnumbered]{algorithm2e}

\newtheorem{lemma}{Lemma}
\usepackage{filecontents}

\begin{document}

\date{}

\title{Geminet: Learning the Duality-based Topology-Agnostic Update Operator for Lightweight Traffic Engineering in Changing Topologies}
\author{
Ximeng Liu\textsuperscript{1,2},
Zhuoran Liu\textsuperscript{1},
Yingming Mao\textsuperscript{3,4},
Yatao Li\textsuperscript{2,5},
Shizhen Zhao\textsuperscript{1},
Xibing Wang\textsuperscript{1}\\[0.3em]
\textsuperscript{1}Shanghai Jiao Tong University \quad
\textsuperscript{2}Zhonggancun Academy \\
\textsuperscript{3}Xi'an Jiaotong University \quad
\textsuperscript{4}Shanghai Innovation Institute \\
\textsuperscript{5}Zhongguancun Institute of Artificial Intelligence
}
\pagenumbering{gobble}
\maketitle
\begin{abstract}
Recently, researchers have explored ML-based Traffic Engineering (TE), leveraging neural networks to solve TE problems traditionally addressed by optimization. However, existing ML-based TE schemes remain impractical: they either fail to handle topology changes or suffer from poor scalability due to excessive computational and memory overhead. To overcome these limitations, we propose Geminet, a lightweight and scalable ML-based TE framework that can handle changing topologies. Geminet is built upon two key insights: (i) decoupling neural networks from topology by learning a topology-agnostic update operator inspired by classical iterative optimization methods (e.g., gradient descent), which depend only on a few gradient-related quantities; (ii) shifting optimization from path-level routing weights to edge-level dual variables, reducing memory consumption by leveraging the fact that edges are far fewer than paths. Evaluations on WAN and data center datasets show that Geminet significantly improves scalability. Its neural network size is only $0.04\%-7\%$ of existing schemes, while handling topology variations as effectively as HARP, a state-of-the-art ML-based TE approach, without performance degradation. When trained on large-scale topologies, Geminet consumes less than $10$ GiB of memory compared to more than $80$ GiB required by HARP, while achieving $18\times$ faster convergence, demonstrating its potential for large-scale deployment.
\end{abstract}
\section{Introduction}
Both data center networks \cite{poutievski2022jupiter,alizadeh2014conga,benson2011microte,al2010hedera,teh2020couder,cao2021trod,chen2018auto} and Wide Area Networks \cite{jiang2009cooperative,kandula2005walking,kandula2014calendaring,kumar2018semi,elwalid2001mate,fortz2000internet,zhong2021arrow,wang2006cope,roughan2003traffic,applegate2003making,suchara2011network,xu2023teal} increasingly rely on traffic engineering (TE) to optimize network performance. TE, typically enabled by a centralized controller \cite{jain2013b4,poutievski2022jupiter,akyildiz2014roadmap, agarwal2013traffic,zaicu2021helix,bogle2019teavar,hong2013achieving,kumar2015bwe,liu2014traffic}, solves optimization problems to efficiently route traffic across network paths. 

Conventional TE schemes typically rely on optimization techniques, such as linear programming. However, as the scale of network topologies grows, conventional optimization methods struggle to swiftly adapt to traffic changes at fine time scales. Consequently, recent research has explored alternative approaches to design TE schemes based on neural network models \cite{alqiam2024transferable,xu2023teal,perry2023dote,liu2024figret}. This shift is driven by the fact that neural networks not only infer significantly faster than conventional models \cite{xu2023teal}, but also have the potential to perform prediction and optimization jointly for better network performance \cite{perry2023dote} and handle traffic bursts \cite{liu2024figret}.

Despite their promising potential, existing ML-based TE schemes face two critical challenges that prevent their practical deployment in production environments:
(i) \textit{Handling variations in topology.} Network topologies continually change due to evolution, failures, or planned maintenance \cite{alqiam2024transferable}. For ML-based TE schemes that cannot handle variations in topology, each change necessitates retraining, which increases operational costs and delays decision-making. Most ML-based TE schemes, such as DOTE \cite{perry2023dote}, FIGRET \cite{liu2024figret}, and TEAL \cite{xu2023teal}, are not explicitly designed for such variations in topology, which limits their practical applicability. (ii) \textit{Scalability in computational resource consumption and memory usage.} While some approaches, such as HARP \cite{alqiam2024transferable} and SaTE \cite{wu2025sate}, are designed to handle changing topologies, they face significant scalability challenges. As network size increases, the hardware requirements of this method grow super-linearly, resulting in prohibitive memory consumption and extremely long training times. To make things worse, the recurrent adjustment unit in HARP requires frequent synchronization, making multi-GPU deployment costly, while SaTE also notes that efficient training of its graph neural networks in multi-GPU environments remains an open challenge.
This discourages network operators from adopting such approaches in production environments.

The reasons why there are no TE schemes that can handle changing topologies while maintaining high scalability can be summarized as follows: 
(i) The lack of a principled methodology for handling changing topologies forces existing approaches to rely on stacking neural network modules. However, these modules may generate a large amount of intermediate data during computation.
(ii) TE requires computing split ratios for each path. However, in large-scale topologies, the number of paths is extremely large, as the number of source-destination pairs grows quadratically with the number of nodes, and each pair may have multiple paths. In existing ML-based TE schemes, these split ratios are generated using neural networks, but the sheer number of paths either results in overly large models or an excessive number of intermediate variables, limiting scalability.

To tackle these challenges, we design Geminet, a lightweight and scalable ML-based TE scheme capable of handling changing topologies without compromising TE performance. Our key idea is to learn a topology-agnostic update operator rather than directly solving the entire TE problem. Intuitively, classical iterative optimization methods, such as gradient descent, update path split ratios using derivatives of the objective function, and these update rules are inherently independent of specific network topologies. Building on this insight, we abstract the update step as a learnable operator that maps gradient-related quantities—such as link utilization and traffic demand—to adjustments of split ratios. Since this mapping remains invariant across topologies, the neural network is fully decoupled from explicit topology inputs. Consequently, Geminet generalizes naturally to diverse and evolving topologies without requiring additional modules to process topology structure.


Through learning the Topology-agnostic update operator, Geminet can handle changing topologies without requiring additional modules. However, as mentioned earlier, when the topology grows, the number of paths increases significantly. If the model iterates over each path, it leads to excessive per-path computations, resulting in high memory and computation overhead. To address this, we harness duality theory and find that determining the optimal values of edge-based dual variables allows us to effortlessly obtain the solution to the original TE problem, i.e., the split ratios on paths. This enables a seamless transition from path-based to edge-based optimization.
Theoretically, this approach transforms the iteration complexity from $O(KN^2)$ at the path level to just $O(N^2)$ at the edge level, where $K$ is the number of paths per source-destination pair and $N$ is the number of nodes. Even more favorably, since the total number of paths is always $KN^2$, open-source network topologies reveal that the actual number of edges is often far smaller than $N^2$, making this approach even more efficient in practice.

We evaluate Geminet using publicly available WAN datasets, as well as data center PoD-level and ToR-level topologies and traffic data. Our key results are as follows: (i) Geminet’s ability to handle all topology variations is on par with HARP without any performance compromises. This holds for any scenarios, including the addition and removal of nodes and links, changes in edge nodes, variations in link capacity (including link failures), and recomputation of tunnels. (ii) Geminet requires significantly fewer parameters, utilizing just 7\% of TEAL’s, 2.6\% of HARP’s, 1.1\% of SaTE's, and 0.04\% of DOTE’s. (iii) In large-scale topologies, where HARP’s training memory consumption exceeds 80 GiB, Geminet requires less than 10 GiB. The convergence speed is $18\times$ faster than HARP's. Additionally, Geminet’s inference time is $2.3\times$ faster than HARP’s. Our code is available at \cite{geminet2026}.

\noindent\textit{This work does not raise any ethical concerns.}

\section{Background and motivation}
\label{sec: background and motivaion}
\subsection{Background: ML-based TE}
\label{sec: background}
\begin{figure*}[ht]
    \centering
    \subfigure[DOTE/FIGRET.]{
    \label{fig: illustration_of_dote}
    \includegraphics[scale=0.42]{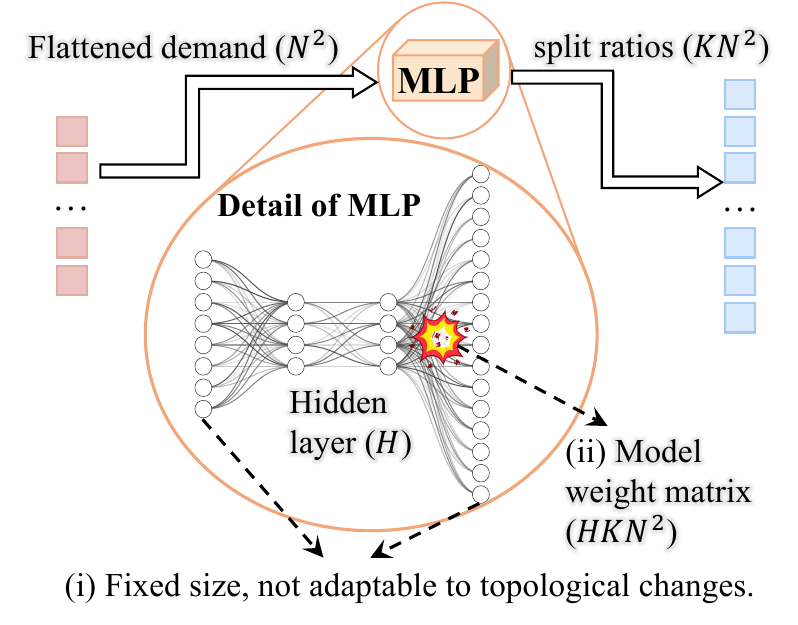}
    }
    \hspace{4mm}
    \subfigure[TEAL. $r_{s_i t_j k}$ represents the split ratio for the $k$-th path from source $i$ to destination $j$.]{
    \label{fig: illustration_of_teal}
    \includegraphics[scale=0.42]{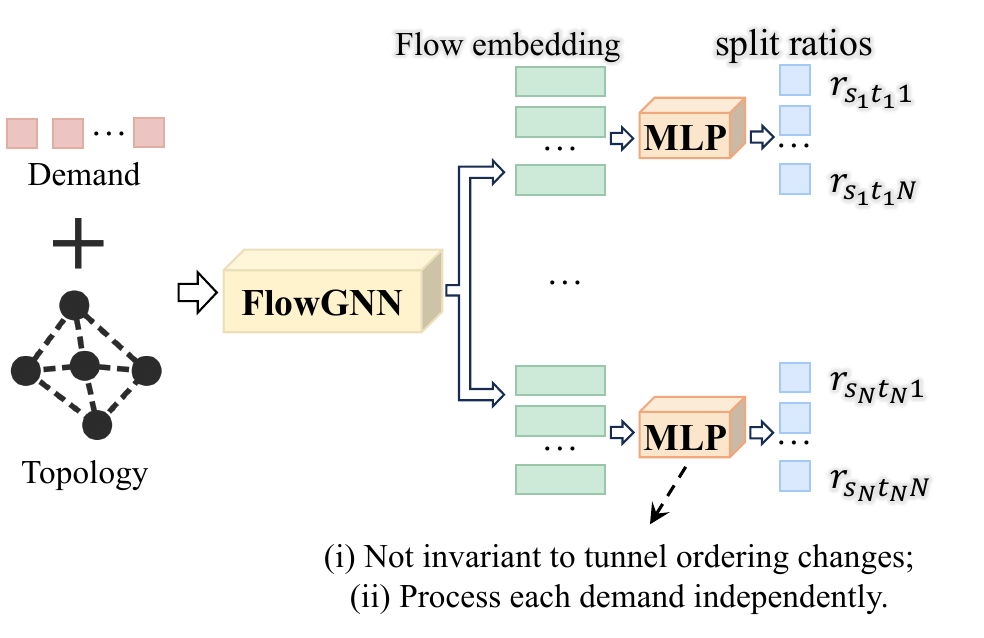}
    }
    
    \subfigure[HARP. The ‘edge embedding’ is generated by a GNN using the topology as input, but this process is omitted for clarity.]{
    \label{fig: illustration of harp}
    \includegraphics[scale=0.43]{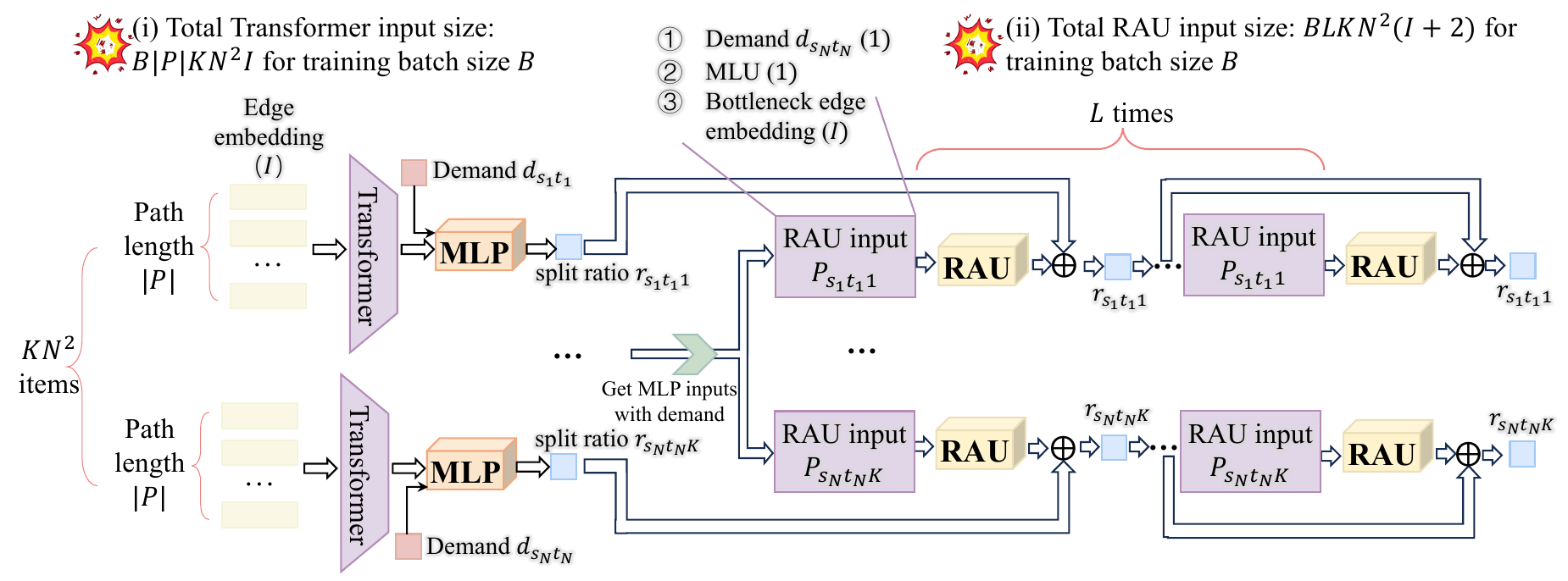}
    }
    \caption{Illustration of existing ML-based TE schemes. Dimensions of key quantities are indicated in parentheses.}
    \label{fig: illustration_of_ML_based_TE}
\end{figure*}

Given a network topology, a traffic matrix, and a set of paths, TE schemes determine traffic routing on each path to optimize a desired network performance objective, such as minimizing the Maximum Link Utilization (MLU). Conventional TE systems are primarily based on linear programming algorithms.

Due to linear programming algorithms struggling to scale solution times with growing network sizes, researchers have recently begun to explore ML-based TE schemes. We introduce several representative ML-based TE schemes as follows:

\textbf{DOTE/FIGRET} \cite{liu2024figret,perry2023dote}: As illustrated in Figure \ref{fig: illustration_of_dote}, both DOTE and FIGRET employ a simple multilayer perceptron (MLP) which receives a flattened traffic matrix as input and directly outputs the flattened split ratios. It has two limitations: (i) This approach does not model nodes, edges, link capacities, or paths, assuming these elements to be static. (ii) Additionally, the neural network is required to output the split ratios for all paths; the size of the neural network is directly related to the total number of paths. 

\textbf{TEAL} \cite{xu2023teal}: As illustrated in Figure \ref{fig: illustration_of_teal}, TEAL models the topology using a bipartite graph of edges and paths, processed by FlowGNN. FlowGNN generates path embeddings by combining Graph Neural Network (GNN) layers, which capture edge-path relationships, with MLP layers, which model interactions among paths within the same flow. To achieve scalability, TEAL does not aggregate all path embeddings into a single large network. Instead, it employs a sharded MLP that processes each source–destination pair independently. However, this approach has two key limitations: (i) it assumes a fixed set of paths for each source-destination pair, and (ii) processing source-destination pairs separately rather than concurrently may compromise TE performance.

\textbf{SaTE} \cite{wu2025sate}: In TEAL, the MLP requires all paths between a source–destination pair as input, which limits its ability to generalize under path variations. SaTE overcomes this by employing three attention-enabled GNNs on different graphs: the original topology, an edge–path bipartite graph, and a flow–path bipartite graph. This design captures global topology, path–edge, and flow–path relationships, making it robust to dynamic topologies and path changes; however, it suffers from scalability issues. The bipartite graphs can be extremely large, as the number of flows scales quadratically with the number of nodes, and the number of paths grows by an additional factor k, representing the number of candidate paths per flow. To mitigate this, SaTE prunes source–destination pairs with zero traffic demand and their corresponding paths. However, this strategy is less effective on dense traffic matrices and is orthogonal to the algorithms themselves, since it can also be applied to other methods designed for dynamic topologies, such as HARP and Geminet. Therefore, our analysis focuses on the algorithms rather than scalability-enhancing strategies.

\textbf{HARP} \cite{alqiam2024transferable}: As illustrated in Figure \ref{fig: illustration of harp}, building on TEAL’s approach of processing source-destination pairs independently, which may be sensitive to path order, HARP processes each path independently to handle changing topologies. It uses a GNN to extract edge embeddings, which are input into a set transformer to compute path embeddings. These are combined with source-destination traffic demands to determine split ratios. To enhance TE performance, HARP introduces a recurrent adjustment unit (RAU). However, while effective with variable topologies, HARP demands substantial computational resources and memory due to (i) the computational cost of the set transformer being directly proportional to path lengths, which grow with network scale, and (ii) iterative adjustments for each path, significantly increasing intermediate variables in large-scale networks.

\textbf{RedTE} \cite{gui2024redte}: In addition to the centralized TE methods, there is also a distributed ML-based TE scheme. During inference, each node maintains a dedicated actor model that determines the split ratio based on previous time step local information, such as its own pending traffic and outgoing link capacities. To address the lack of a global perspective, RedTE uses a centralized training framework in which each node has both a local actor and a global critic that evaluates the actor’s output using global topology and traffic information. This allows each node to learn its role within the topology during training and make near-global decisions using only local information during inference. However, this approach has two main limitations: (1) it does not support variable topologies, as each node is associated with a dedicated actor. When the topology changes and node roles shift, the actor models must be retrained; \begin{table*}[t]
\centering
\scalebox{0.75}{
\begin{tabular}{c|cc|cc|cc|cc|cc}
\toprule
\multirow{2}{*}{TE scheme} 
& \multicolumn{2}{c|}{\begin{tabular}[c]{@{}c@{}}GEANT\\ \#Node 22, \#Edge 72\end{tabular}} 
& \multicolumn{2}{c|}{\begin{tabular}[c]{@{}c@{}}Cogentco\\ \#Node 197, \#Edge 486\end{tabular}} 
& \multicolumn{2}{c|}{\begin{tabular}[c]{@{}c@{}}Meta WEB DC\\ \#Node 324, \#Edge 12796\end{tabular}} 
& \multicolumn{2}{c|}{\begin{tabular}[c]{@{}c@{}}KDL\\ \#Node 754, \#Edge 1790\end{tabular}} 
& \multicolumn{2}{c}{\begin{tabular}[c]{@{}c@{}}ASN\\ \#Node 1739, \#Edge 8558\end{tabular}} \\
\cmidrule(lr){2-11}
& Model Mem & Total Mem
& Model Mem & Total Mem
& Model Mem & Total Mem
& Model Mem & Total Mem
& Model Mem & Total Mem \\
\midrule
DOTE/FIGRET   & 0.004  & 0.02 & 0.30   & 0.53 & 0.80   & 1.33 & 4.36   & 9.97  & 7.25   & 40.10 \\
TEAL          & 2.6e-5 & 0.02 & 2.6e-5 & 0.78 & 2.6e-5 & 1.48 & 2.6e-5 & 14.6  & 2.6e-5 & 43.49 \\
SaTE          & 4.2e-5 & 0.19 & 4.2e-5 & 48.88 & 4.2e-5 & >80 & 4.2e-5 & >80 & 4.2e-5 & >80\\
HARP          & 7.3e-5 & 0.07 & 7.3e-5 & 18.8 & 7.3e-5 & 10.62& 7.3e-5 & >80   & 7.3e-5 & >80   \\
\textbf{Ours} & 1.9e-6 & 0.02 & 1.9e-6 & 0.45 & 1.9e-6 & 0.99 & 1.9e-6 & 9.83  & 1.9e-6 & 20.50 \\
\bottomrule
\end{tabular}
}
\caption{Comparison of GPU memory usage for training across different topologies. The data in the Table is in GiB. Ours, i.e., Geminet, is proposed in \S \ref{sec: design} and evaluated in \S \ref{sec: evaluation}.}
\label{tab: memory_usage}
\end{table*}(2) it incurs high training overhead, as each node has its critic that processes global information; as the topology grows, the number and size of critics increase.

\noindent\textbf{Why not use a single neural network to learn all parameters?}
End-to-end designs typically require fixed-dimensional inputs/outputs that scale with the number of OD pairs/paths/edges, making the model topology-specific (e.g., DOTE/FIGRET) and not directly reusable under topology changes.

\noindent\textbf{Why periodic retraining is insufficient.}
One may ask whether a topology-specific neural TE model can be periodically retrained after topology updates. However, topology updates can be frequent in operational WANs (e.g., tens of distinct topology changes within a few weeks as reported by HARP \cite{alqiam2024transferable}). More importantly, for models with fixed-dimensional inputs/outputs (e.g., DOTE/FIGRET), a topology or tunnel update changes the dimensionality and semantics of the model interface, making the existing model unusable immediately. Retraining after each update would therefore introduce a non-trivial window with no usable learned policy, forcing operators to fall back to slower optimization solvers and defeating the goal of learning-based TE in practice.
\subsection{ML-based TE needs streamlining}
\label{sec: ml-based TE needs streamlining}
\noindent\textbf{Poor scalability in terms of computational resource consumption and memory usage.} As observed in \S \ref{sec: background}, the trend in the development of ML-based TE indicates that researchers are increasingly stacking more modules into the architecture to enhance functionality and performance. However, deep learning is not a panacea. While stacking various deep learning modules can enhance performance, it also significantly increases the demand for computational resources. For HARP, the state-of-the-art algorithm that supports variable topologies, training for deployment in large-scale topologies may require hundreds of gigabytes of GPU resources. To illustrate this, we summarize the memory usage of various ML-based TE algorithms in Table \ref{tab: memory_usage}. The reported memory consumption is measured using the peak GPU memory allocation, as tracked by PyTorch’s built-in memory profiling tools. Training is conducted with a batch size of 16 in single-precision. The table presents two types of memory usage: (i) Model Mem, corresponding to the memory associated with neural network parameters, and (ii) Total Mem, representing overall GPU consumption. It can be seen that GPU usage depends not only on model size but also on additional intermediate variables generated during computation. Notably, HARP, when trained on a topology with 754 nodes and 1,790 edges, occupies over 80 GiB of memory, with inference memory demands reaching 71.85 GiB.
We exclude RedTE from the table because its memory consumption exceeds 80 GiB on all but the smallest topologies (e.g., GEANT), making it impractical to profile under consistent settings. Instead, in \S\ref{sec: geminet under traffic uncertainty}, we provide a theoretical analysis of RedTE’s neural network memory usage to illustrate the root cause of its excessive training overhead. It is worth noting that in \S\ref{sec: handling drastically different topology} we evaluate the performance of HARP when applied to topologies that are drastically different from the training topology, and observe sub-optimality. This indicates that training on small topologies and directly deploying on large and different ones is impractical.

\noindent\textbf{Adding more hardware resources is not a cost-effective solution.} First, hardware demand grows super-linearly with network size, making this approach unsustainable. Real-world networks can be much larger than the topology discussed earlier, and as they continue to expand, they make hardware-based scaling a bottomless pit. 
Second, even with sufficient hardware, HARP’s architecture complicates distribution across multiple GPUs. Existing parallelization techniques, such as tensor parallelism (splitting within layers) and pipeline parallelism (splitting between layers) \cite{shoeybi2019megatron}, become inefficient under HARP’s RAU process. Since the RAU requires numerous iterative updates, distributing it across different GPUs necessitates frequent data exchanges among them, leading to significant communication overhead. Consequently, distributing HARP across GPUs is impractical. In contrast, our approach can be implemented on a single GPU, eliminating costly multi-GPU coordination.

\noindent\textbf{Reasons for poor scalability.} To understand the reasons for poor scalability, we first outline the key sources of memory consumption in neural network training \cite{sohoni2019low}. These include model memory, which stores the neural network’s parameters; optimizer memory, which holds gradients and momentum buffers; and activation memory, which stores intermediate values required during backpropagation, including the outputs of each layer in the forward pass and their corresponding gradients in the backward pass. Thus, memory consumption is influenced not only by model size but also by the intermediate variables generated during computation. Based on this, the scalability limitations of existing approaches can be attributed to two main factors: (i) Stacking modules to handle changing topologies. Methods using graph neural networks for topology modeling or transformers for path embeddings introduce considerable computational overhead. For instance, self-attention in transformers processes each path individually, generating intermediate variables that scale quadratically with path length (see Appendix \ref{sec: self-attention} for details). (ii) The burden of optimizing large path outputs. As the network size increases, the number of paths grows significantly. A network with $N$ nodes contains $N(N-1)$ source-destination pairs, each with $K$ available paths, resulting in a total of $KN(N-1)$ paths. Schemes like DOTE, which directly output TE decisions for all paths, require large models due to high-dimensional outputs, while HARP generates excessive intermediate variables due to the need for per-path adjustments.

\subsection{Insights to improve scalability}
\noindent\textbf{Decoupling topology-related parameters from neural networks.} Given the computational and memory overhead of handling topology within neural networks, we ask: can topology be decoupled from the neural network? Fortunately, the gradient descent update rule is inherently topology-agnostic, relying only on a few gradient-related quantities—such as link utilization and traffic demand—rather than explicit topology structures. This leads to our insight: instead of solving TE decisions directly, we use the neural network to learn the iterative adjustment process, mapping these gradient-related quantities to split ratio updates.

\noindent\textbf{Leveraging duality to shift from path-based to edge-based optimization.} The total number of paths in a network is large, making direct path-level TE decisions memory-intensive. However, as shown in Table \ref{tab: memory_usage}, the number of edges is significantly smaller. Even in the extreme case of a fully connected graph, the number of edges remains only $\frac{1}{K}$ of the total number of paths, where $K$ is the number of paths per source-destination pair. This observation motivates us to explore whether duality theory can transform path-level TE optimization into solving dual variables associated with edges, significantly reducing memory consumption.

\section{Model}
\label{sec: model}
\textbf{Notations \& Definitions.} We introduce recurring mathematical notations and definitions of TE. All notations used in this paper are also tabulated in Table \ref{tab: notation table} for ease of reference.
\begin{itemize}[leftmargin=*, topsep=0pt, partopsep=0pt, itemsep=0pt]
    \item \textbf{Network.} Network topology is represented as a graph $G(V,E,c)$, where $V$ and $E$ are the node and edge sets, respectively, and $c:E \rightarrow \mathbf{R}^+$ assigns capacities to edges.
    \item \textbf{Traffic demands.} A demand matrix $D$ is of size $|V| \times |V|$, where $D_{st}$ represents the traffic demand from $s$ to $t$.
   \item \textbf{Paths.} Each source $s$ communicates with each destination $t$ via a set of network paths $P_{st}$.
   \item \textbf{TE configurations.} A TE configuration $\mathcal{R}$ specifies for each path $p \in P_{st}$ a split ratio $r_p$, where $r_p$ is the fraction of the traffic demand from $s$ to $t$ forwarded along path $p$. The split ratios must satisfy $\sum_{p\in P_{st}}r_p=1$.
   \item \textbf{TE objective.} We focus on Maximum Link Utilization (MLU), a widely used TE metric \cite{azar2003optimal,benson2011microte,chiesa2016lying,poutievski2022jupiter,valadarsky2017learning}, for concreteness. 
   Given a demand matrix $D$ and TE configuration $\mathcal{R}$, the total traffic traversing an edge $e$ is $f_e=\sum_{s,t\in V}\sum_{p\in P_{st}, p\owns e} D_{st} r_p$, and then the MLU induced by $D$ and $\mathcal{R}$ is $\max_{e\in E}\frac{f_e}{c(e)}$. 
\end{itemize}
\textbf{TE model.} Given the above definitions, Traffic Engineering optimization can be formulated in the form of Equation (\ref{equ: te opt formulation}):
\begin{equation}
\boxed{
\label{equ: te opt formulation}
\begin{alignedat}{2}
&\mathop{\text{minimize}}\limits_{\mathcal{R},m} &&\quad \quad m \\
&\text{subject to} \ 
&&\sum_{s,t\in V}\sum_{p\in P_{st},p \owns e} D_{st} \cdot r_p \leq m \cdot c(e), \quad \forall e \in E, \\
& &&\sum_{p \in P_{st}} r_p = 1, \quad \forall s, t \in V, \\
& && r_p \geq 0, \quad \forall s,t \in V, p\in P_{st}.
\end{alignedat}
}
\end{equation}
\section{Design}
\label{sec: design}
\subsection{Topology-agnostic update operator}
\label{sec: learning the iterative process}
In TE, changes in network topology and paths pose significant challenges for learning-based methods. A common approach is to feed raw topology/path structures into neural models, which increases complexity and makes cross-topology transfer more data-hungry and sensitive to distribution shifts in practice. To address this issue, we propose a \textbf{\textit{topology-agnostic update operator}}. Here, \emph{topology-agnostic} means that the \emph{learned update operator itself} is agnostic to \emph{learned topology embeddings}: its parameters and input/output interface remain unchanged across topologies. Importantly, topology is still used in an analytic and pre-computable manner when computing objective/gradient-related quantities such as flows and MLU (e.g., through the path-to-edge incidence matrix and link capacities). Our method compresses these topology-dependent computations into a small set of computable, gradient-related features, which are then fed into a lightweight neural network. In this way, the update mechanism can be reused across arbitrary topologies.

\noindent\textbf{Generalized update operator formulation.} Concretely, let $M(\mathcal{R})$ denote the objective in Equation (\ref{equ: te opt formulation}), and assume an initial solution $\mathcal{R}$ is given. The update of $\mathcal{R}$ towards the optimum can be expressed in a generalized form as:
\begingroup
\setlength{\abovedisplayskip}{3pt}
\setlength{\belowdisplayskip}{3pt}
\setlength{\abovedisplayshortskip}{3pt}
\setlength{\belowdisplayshortskip}{3pt}
\begin{equation}
r_p^{\text{new}} = r_p - \mathcal{G}(s_p),
\label{equation: generalized update}
\end{equation}
\endgroup
where, $\mathcal G(\cdot)$ denotes the update operator, and $s_p$ is the state feature vector of path $p$. This unified form covers a broad range of iterative optimization methods, such as gradient descent, Newton’s method, and projected subgradient methods \cite{boyd2004convex}. In other words, the existence of the update operator is not an extra assumption but rather is naturally ensured by the construction of these standard methods.
Notably, HARP's RAU can be viewed as a concrete instantiation of the generalized update operator in Equation~(\ref{equation: generalized update}).
Here we make this connection explicit through a mathematical interpretation, and highlight that the recurrent update operator---rather than learned topology encoders---is the key mechanism for enabling cross-topology reuse under changing topologies.

\noindent\textbf{Design of path state vector.} In principle, the state vector $s_p$ must capture the local sensitivity of the objective function with respect to $r_p$, as any update rule fundamentally relies on gradient- or curvature-related information. This aligns with classical optimization: gradient descent uses first-order derivatives, Newton’s method incorporates second-order terms, and subgradient methods rely on generalized gradients. Thus, $s_p$ can be viewed as a compact representation of such derivative information. In the TE problem, however, $M(\mathcal{R})$ is piecewise linear, so higher-order derivatives are identically zero within linear regions or are undefined at non-differentiable boundaries. Consequently, the update reduces to relying solely on first-order information.

From the definition of $M(\mathcal{R})$, the first-order derivative with respect to $r_p$ is given by:
$\frac{\partial M(\mathcal{R})}{\partial r_p} = \frac{D_{st}}{c(e)}$ if the path $p$ includes a link  $e$ whose \( \text{Utilization}(e) = \frac{\sum_{s,t \in V} \sum_{p \in P_{st}, p \owns e} D_{st} \cdot r_p}{c(e)} \) equals the MLU $m$; otherwise, $\frac{\partial M(\mathcal{R})}{\partial r_p} = 0$. 
Therefore, the gradient value for $r_p$ is completely determined by the following four quantities: (i) the utilization of the most heavily utilized link on the path $U(e_{\text{heavy}})$, (ii) the global maximum link utilization MLU, (iii) the traffic demand flowing through the path $D_{st}$, and (iv) the capacity of the most heavily utilized link $c(e_\text{heavy})$. We directly collect them into the path state vector $s_p$.
Note that $U(e_{\text{heavy}})$ and $\mathrm{MLU}$ are computed deterministically from the current topology, link capacities, traffic demands, and split ratios; thus, topology changes affect only the \emph{values} of these features, not the parameterization of the learned mapping.

\noindent\textbf{Neural network realization of the update operator.} With the above, the update rule can be rewritten in the form of Equation (\ref{equ: gradient descent detail}). Here, $f(\cdot)$ serves as the concrete realization of the operator $\mathcal{G}(\cdot)$, tailored to the first-order derivative.
\begingroup
\setlength{\abovedisplayskip}{3pt}
\setlength{\belowdisplayskip}{3pt}
\setlength{\abovedisplayshortskip}{3pt}
\setlength{\belowdisplayshortskip}{3pt}
\begin{equation}
    r_p^{\text{new}}=r_p-f\left(U(e_{\text{heavy}}),\text{MLU},D_{st},c(e_{\text{heavy}})\right)
\label{equ: gradient descent detail}
\end{equation}
\endgroup

We use a neural network to learn the function $f$, which takes the four quantities $U(e_{\text{heavy}}), \text{MLU}, D_{st}, c(e_{\text{heavy}})$ as input and outputs the adjustment for $r_p$. 
As discussed above, $f$ is topology-agnostic in the sense that its parameterization and input/output interface are fixed and do not depend on any learned topology embeddings; hence, the neural network is decoupled from topology-specific encoders and can be reused across topologies.
Owing to the low dimensionality of both the input and the output, a lightweight MLP is sufficient to learn $f$. 
Notably, the inputs $U(e_{\text{heavy}})$, $\mathrm{MLU}$, and $c(e_{\text{heavy}})$ encapsulate topology-related information that can be computed with simple calculations; hence, changes in topology are reflected through these features, while the learned mapping $f$ itself remains independent of any particular topology and generalizes across topologies.

\begin{figure}[!ht]
    \centering
    \includegraphics[scale=0.84]{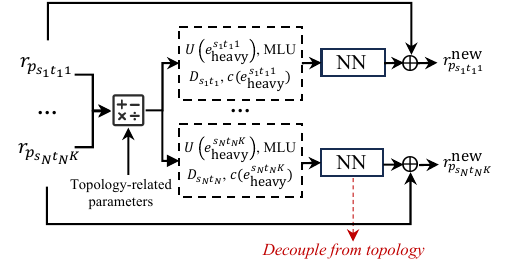}
    \caption{Illustrating the topology-agnostic update operator: The neural network (NN) is decoupled from topology, with topology information fused into variables through simple mathematical operations before input to the NN.}
    \label{fig: illustration_of_iterative}
\end{figure}

The design of the topology-agnostic update operator, how it decouples the neural network from topology information, and how topology information is incorporated through simple computations is illustrated in Figure \ref{fig: illustration_of_iterative}.


\subsection{Harnessing duality}
\label{sec: harness duality in TE}
As mentioned, the update operator can be applied to either the primal variables or the dual variables. Considering that the number of primal variables $r_p$ in the TE problem is $KN(N-1)$, where $N$ is the number of nodes, applying the update operator over the primal variables would result in a large number of intermediate variables, leading to higher memory usage and increased computational resource consumption. Therefore, in this section, we leverage duality theory to transform the updates on the primal variables into updates on the edge-associated dual variables. The maximum number of edges is $N(N-1)$, and as shown in the topology details in Table \ref{tab: memory_usage}, the actual number of edges is often much smaller, significantly reducing the intermediate variable count and thereby decreasing memory usage.

\noindent\textbf{Formulate the Lagrangian.} For the optimization problem defined by Equation (\ref{equ: te opt formulation}), the constraints $\sum_{p \in P_{st}} r_p = 1, \forall s,t \in V$ and $r_p \geq 0, \forall s,t \in V, p \in P_{st}$ can be easily satisfied through normalization \cite{liu2024figret}. Consequently, we apply a partial Lagrangian solely to the first constraint. To do this, we introduce a Lagrangian multiplier vector $\lambda$ (where $\lambda \in \mathbb{R}^{|E|}$) and formulate the Lagrangian as Equation (\ref{equ: lagrangian partial}). At this stage, the feasible region \(\mathcal{R}_{\text{feas}}\) for \(\mathcal{R}\), defined by \(
\mathcal{R}_{\text{feas}} = \{ \mathcal{R} \in \mathbb{R}^{KN^2} \mid \sum_{p \in P_{st}} r_p = 1, \forall s, t \in V, \text{ and } r_p \geq 0, \forall s, t \in V, p \in P_{st} \}.
\)
\begingroup
\setlength{\abovedisplayskip}{3pt}
\setlength{\belowdisplayskip}{3pt}
\setlength{\abovedisplayshortskip}{3pt}
\setlength{\belowdisplayshortskip}{3pt}
\begin{equation}
\begin{aligned}
L(\mathcal{R},m,\lambda) &= m + \sum_{e\in E}\lambda_e (\sum_{s,t\in V}\sum_{p\in P_{st},p \owns e}D_{st}r_p-m \cdot c(e))\\
&= \bigg(1-\sum_{e\in E}\lambda_e c(e)\bigg)m + \sum_{e\in E}\sum_{s,t\in V}\sum_{p\in P_{st},p \owns e}\lambda_e D_{st}r_p\\ 
&= \bigg(1-\sum_{e\in E}\lambda_e c(e)\bigg)m + \sum_{s,t\in V}\sum_{p\in P_{st}}\sum_{e\in p} \lambda_e D_{st}r_p. 
\end{aligned}
\label{equ: lagrangian partial}
\end{equation}
\endgroup

The transition from the second to the third line in Equation (\ref{equ: lagrangian partial}) involves swapping the order of summation. A step-by-step derivation is provided in Appendix \ref{sec: summation order exchange detail} for completeness.

Dualizing only the capacity constraints while keeping the simplex constraints in the primal domain is standard and yields the same LP dual as the full Lagrangian construction, since the dual function is defined by minimizing $L(\mathcal{R},m,\lambda)$ over $\mathcal{R}\in \mathcal{R}_{\text{feas}}$.
Because the underlying TE formulation is a feasible and bounded linear program, strong duality holds.

\noindent\textbf{Derive the primal optimal from the Lagrangian.} Since the primal problem is a linear programming problem, the optimal solution for $\mathcal{R}$ can be directly obtained by optimizing the Lagrangian function, as shown in Equation (\ref{equ: te lagrangian with optimal dual}), where $\lambda^*$ is the optimal solution of $\lambda$. 
\begingroup
\setlength{\abovedisplayskip}{3pt}
\setlength{\belowdisplayskip}{3pt}
\setlength{\abovedisplayshortskip}{3pt}
\setlength{\belowdisplayshortskip}{3pt}
\begin{equation}
    \mathop{\text{minimize}}_{\mathcal{R},m}\bigg\{L(\mathcal{R},m,\lambda^*)|\mathcal{R}_{\text{feas}} \bigg\}.
\label{equ: te lagrangian with optimal dual}
\end{equation}
\endgroup
Since only the second term involves $\mathcal{R}$ in Equation (\ref{equ: lagrangian partial}), it can be rewritten in the form of Equation (\ref{equ: revised equation}).
\begingroup
\setlength{\abovedisplayskip}{3pt}
\setlength{\belowdisplayskip}{3pt}
\setlength{\abovedisplayshortskip}{3pt}
\setlength{\belowdisplayshortskip}{3pt}
\begin{equation}
    \mathop{\text{minimize}}_{\mathcal{R},m}\bigg\{\sum_{s,t\in V}\sum_{p\in P_{st}}\sum_{e\in p} \lambda_e^* D_{st}r_p|\mathcal{R}\in \mathcal{R}_{\text{feas}}\bigg\}.
\label{equ: revised equation}
\end{equation}
\endgroup

Furthermore, the optimization problem specified in Equation (\ref{equ: revised equation}) can be rewritten as $\mathop{\text{minimize}}_{\mathcal{R},m}\bigg\{\sum_{s,t\in V}D_{st}\sum_{p\in P_{st}}(\sum_{e\in p}\lambda_e^*)r_p|\mathcal{R}\in \mathcal{R_{\text{feas}}}\bigg\}$. This reformulation shows that the objective function and the feasible region constraint are \textit{independent} for each source-destination pair $(s,t)$. Given this structuring, we can decouple the optimization for each source-destination pair under the premise of maintaining optimality. The optimization problem for each pair can be formulated as in Equation (\ref{equ: formulate for each pair}).
\begin{equation}
\boxed{
\begin{aligned}
    &\mathop{\text{minimize}}\limits_{r}\qquad &&\sum_{p\in P_{st}}D_{st}(\sum_{e\in p} \lambda_e^*)r_p \\
    &\text{subject to}&&\sum_{p\in P_{st}} r_p=1, r_p \geq 0, \forall p\in P_{st}.
\end{aligned}
}
\label{equ: formulate for each pair}
\end{equation}
The solution to Equation (\ref{equ: formulate for each pair}) is remarkably simple. Since the $D_{st}$ can be treated as a constant, it can be temporarily set aside. 
The objective $\sum_{p \in P_{st}} (\sum_{e \in p} \lambda_e^* )r_p$ is minimized by placing all the weight on a path $p \in P_{st}$ with the smallest $\sum_{e \in p} \lambda_e^*$ (i.e., the ``shortest'' path under edge weights $\lambda^*$). 
If the minimum-cost path is unique, the optimal solution is one-hot: assign $r_p = 1$ to that path and $r_p = 0$ to all others. 
If there are ties (i.e., multiple paths attain the same minimum $\sum_{e \in p} \lambda_e^*$), the primal optimum is not unique: any split-ratio vector that assigns total mass $1$ over the tied minimum-cost paths is optimal.

In deep learning training, we implement this mapping by applying the Softmin operation to $\sum_{e \in p} \lambda_e^*$ across all paths $p \in P_{st}$. Softmin transforms these values into a probability distribution and handles ties deterministically: equal-cost paths receive equal probabilities, while ensuring that the probabilities sum to 1. Compared to hard minimization, this yields a smooth, differentiable approximation; with a finite smoothing strength, it may assign a small probability mass to near-minimum-cost paths, trading exact hard-min selection for differentiability and numerical stability.

\noindent\textbf{Update operator on dual variables.}
By now, we know that as long as the optimal dual variables can be obtained, the optimal splitting ratio can be easily determined. The next question is how to adjust the dual variables by defining an update operator on the dual space, analogous to the operator introduced in \S \ref{sec: learning the iterative process} for the primal variables. To address this, we first compute the derivative of Equation (\ref{equ: lagrangian partial}) to the dual variable. For any $\lambda_e$, the derivative is given in Equation (\ref{equ: partial e of lagrangian}).
\begingroup
\setlength{\abovedisplayskip}{3pt}
\setlength{\belowdisplayskip}{3pt}
\setlength{\abovedisplayshortskip}{3pt}
\setlength{\belowdisplayshortskip}{3pt}
\begin{equation}
\frac{\partial L}{\partial \lambda_e} = \sum_{s,t \in V} \sum_{p \in P_{st}, p \owns e} D_{st} r_p - m \cdot c(e).
\label{equ: partial e of lagrangian}
\end{equation}
\endgroup
\begingroup
\setlength{\abovedisplayskip}{3pt}
\setlength{\belowdisplayskip}{3pt}
\setlength{\abovedisplayshortskip}{3pt}
\setlength{\belowdisplayshortskip}{3pt}
\begin{equation}
    \lambda_e^{\text{new}}=\lambda_e - f_{\text{Dual}}(F(e),\text{MLU},c(e)).
\label{equ: f_dual}
\end{equation}
\endgroup
Based on Equation (\ref{equ: partial e of lagrangian}), $f_\text{Dual}(\cdot)$ serves as the concrete realization of the update operator on the dual variables. It takes three inputs—(i) the flow passing through the edge $F(e)$, (ii) the MLU, and (iii) the edge capacity $c(e)$—and produces one output, the adjustment $\Delta \lambda$. We learn this mapping using a neural network. The adjustment procedure for $\lambda$ is as follows: given $\lambda$, we first compute the corresponding $\mathcal{R}$ using summation and Softmin operations. Based on $\mathcal{R}$, we derive the three input features and feed them into the neural network, which outputs the update for $\lambda_e$. This process is repeated until a predefined number of iterations is reached. The procedure is summarized in Pseudocode \ref{algorithm: Pseudocode for adjust dual variable} in Appendix \ref{sec: Pseudocode for adjust dual variable}.
\subsection{Framework design}
\label{sec: framework design}
\noindent\textbf{Initial dual variable generation module design.} We first design how to solve for the initial dual variable. In principle, the update operator can be applied starting from any value; however, a well-chosen initialization can reduce the iterations required. The selection method is not unique, and we adopt a heuristic approach.
As can be seen from Equation (\ref{equ: formulate for each pair}), $\lambda$ can be interpreted physically as the "congestion price". When the "congestion price" on a path ($\sum_{e\in p}\lambda_e$) is too high, $r_p$ should be smaller to avoid excessive network congestion. Therefore, the overarching idea for determining the initial value is that this value should reflect a certain degree of congestion.

Our approach is as follows: For each path, we use a neural network that receives the traffic flow for the source-destination pair it serves, along with the bottleneck edge capacity ($\min_{e\in p} c(e)$) of that path. It outputs a potential load value for each path, forming a potential load vector $\Psi^{(1,|\Phi|)}$, where $\Phi$ denotes the set of all paths. Subsequently, the potential loads of the paths passing through each edge are aggregated to determine the “potential load” for each edge, which can be computed by calculating $\Psi^{(1,|\Phi|)} \times \xi^{(|\Phi|,|E|)}$, where $\xi^{(|\Phi|,|E|)}$ is the path-to-edge matrix, and $\xi_{i,j}=1$ indicates that path $i$ contains edge $j$.

\begin{figure}[!ht]
    \centering
    \includegraphics[scale=0.45]{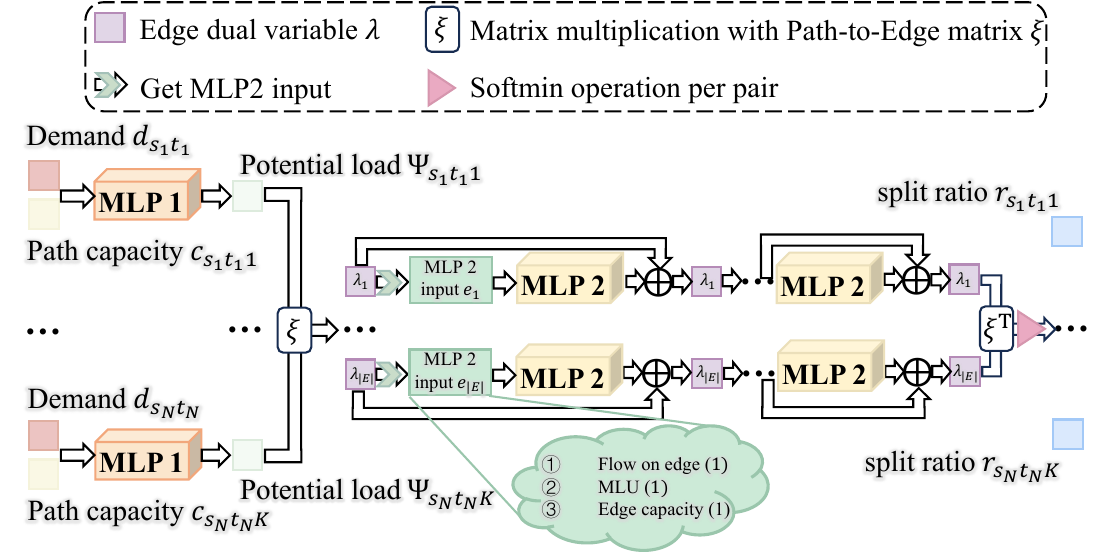}
    \caption{Illustration of Geminet’s overall framework, which can be divided into two parts: the Initial Dual Variable Generation Module and the Update Module. MLP~1 predicts the potential load for initializing the dual variables; MLP~2 implements the dual-update operator described in \S\ref{sec: harness duality in TE}. Blocks with the same label (e.g., MLP~1, MLP~2) share parameters.}
    \label{fig: illustration_of_dualte}
\end{figure}

\noindent\textbf{Overall framework of Geminet.}
The overall framework of Geminet comprises two main components: the Initial Dual Variable Generation Module and the Update Module, as illustrated in Figure \ref{fig: illustration_of_dualte}.
The workflow proceeds as follows. First, for each path, the demand it serves and the capacity of its bottleneck edge are used as inputs to the MLP 1, which generates the potential path load. Based on the mapping between paths and edges, the potential path loads are then aggregated to obtain the potential edge loads, serving as the initial values of the dual variables. Then, the dual update operator (MLP 2) iteratively refines the dual variables for a limited number of iterations. Finally, the updated dual variables are used to compute the split ratios-the output of Geminet.

\noindent\textbf{Training methodology.} Geminet is trained end-to-end, where the entire workflow-from the initial demand and bottleneck capacity inputs to the final split ratios-is optimized jointly. This design allows all components to be trained simultaneously under a unified objective, eliminating the need for separate optimization stages. Such an end-to-end paradigm has proven to be effective in the machine learning community \cite{jumper2021highly}.

\section{Evaluation}
\label{sec: evaluation}
\subsection{Methodology}
\label{sec: methodology}
 We evaluate Geminet using real-world datasets and compare Geminet with state-of-the-art ML-based traffic engineering schemes. For detailed information about the evaluation setup, including machine specifications, CUDA version, and other configurations, please refer to Appendix \ref{sec: evaluation setup}.\\
\textbf{Topologies.} We evaluate Geminet on both single-cluster and multi-cluster topologies. Following \cite{alqiam2024transferable}, a cluster is defined as a group of topologies sharing the same set of nodes and edges. A topology is single-cluster if its nodes and edges remain unchanged across all snapshots; otherwise, it is multi-cluster.

For single-cluster topologies, we cover three settings where TE is commonly applied—WANs, the AS-level Internet, and DCNs—using canonical datasets from prior work~\cite{liu2024figret,poutievski2022jupiter,xu2023teal}. 
For WANs, we use GEANT~\cite{uhlig2006providing} and two Topology Zoo networks (Cogentco and KDL)~\cite{knight2011internet}. 
For the AS-level Internet, we use the CAIDA ASN topology~\cite{caida-asn}. 
For DCNs, we use the Meta WEB DC~\cite{roy2015inside} and model both PoD and ToR levels with a direct-connect topology~\cite{liu2024figret}: PoD is fully connected, and ToR is a random regular graph.
For multi-cluster topologies, we evaluate on DynGeant, a dynamic variant of GEANT that evolves over time through changes in topology and link capacities. Since the HARP dataset is not publicly available, we construct DynGeant with dynamic characteristics similar to those described in HARP \cite{alqiam2024transferable}. The topological variations of DynGeant are visualized in Appendix \ref{section: details of dyngeant}, specifically in Figure \ref{fig: variations across clusters} and Figure \ref{fig: variations within clusters}.

For each cluster, we use Yen’s algorithm to precompute the 4 shortest \cite{kumar2018semi} paths between each source-destination pair, serving as the candidate paths for flow allocation.

\noindent\textbf{Traffic data.} For GEANT and Meta, we use publicly available traffic traces~\cite{uhlig2006providing,roy2015inside}. 
For Cogentco, KDL (Topology Zoo), and ASN, we generate traffic using the gravity model~\cite{applegate2003making,roughan2002experience}. 
For DynGeant, we use the GEANT trace and apply the gravity model only for newly introduced source--destination pairs due to node additions. 
Unless otherwise specified, we use 75\% of the data for training and 25\% for testing.

\noindent\textbf{Metrics.} 
(i) TE Performance: The quality of the TE decision. It is presented by the Normalized MLU, defined as the ratio of the MLU achieved by the ML-driven TE scheme to that of the optimal solution (obtained using the Gurobi optimization solver).
(ii) Memory Usage: The GPU memory consumption during training.
(iii) Time-to-Performance: The training time required to achieve specific performance targets. 

\noindent\textbf{Baselines.}
We compare Geminet with state-of-the-art ML-based TE schemes, including HARP, SaTE, DOTE, TEAL, RedTE, and FIGRET. HARP, SaTE, DOTE, TEAL, and RedTE are described in \S\ref{sec: background}. FIGRET, designed for traffic uncertainty, shares DOTE’s architecture but introduces an additional loss function to enhance robustness. If the input is the exact traffic matrix at the current time step, traffic uncertainty is absent, and FIGRET degenerates to DOTE. RedTE also determines split ratios based on local information from the previous time step. Therefore, both FIGRET and RedTE are evaluated only under the traffic-uncertainty setting (\S\ref{sec: geminet under traffic uncertainty}). 
Hyperparameter details are provided in Appendix \ref{sec: Hyperparameter Search}.
\subsection{Handling topology changes}
\label{sec: handling topology changes}
We compare the performance of Geminet with HARP and SaTE on the changing topology DynGeant. The characteristics of DynGeant and our train/test split are described in \S\ref{sec: methodology}; all results reported below are obtained on the test set. The results evaluated on the test set are summarized in Table \ref{tab: comparing harp geminet sate}. From these results, we make two key observations:
\begin{itemize}[leftmargin=*, topsep=0pt, partopsep=0pt, itemsep=0pt]
    \item Under the premise that Geminet and HARP perform similarly, the memory usage of Geminet is $\mathbf{4\%}$ of that of HARP. Regarding convergence speed, Geminet reaches the target performance $\mathbf{1.9\times}$, $\mathbf{2.7\times}$, and $\mathbf{3.6\times}$ faster at the 1.2, 1.18, and 1.16 Normalized MLU thresholds, respectively.
    \item For HARP, memory consumption for handling similar-scale topologies, GEANT and DynGeant, is 0.07 and 0.53, respectively. This is because HARP first applies a graph neural network to generate edge embeddings and then uses a transformer to derive path embeddings. In static topologies, where the network structure and edge-path mappings remain unchanged, embeddings are computed only once. However, in changing topologies, HARP must recompute embeddings for each data point.
    \item For SaTE, its performance is inferior to that of Geminet and HARP. Methods that mainly rely on GNN and reinforcement learning do not achieve the best results in minimizing MLU, as demonstrated in several studies, including SaTE itself \cite{wu2025sate,liu2024figret,alqiam2024transferable}. The underlying reason for this phenomenon requires further investigation.
\end{itemize}
\begin{table}[!ht]
\centering
\scalebox{0.8}{
\begin{tabular}{cc|ccc}
\toprule
                                                                                                             &           & Geminet & HARP & SaTE \\ \midrule
\multicolumn{1}{c|}{\multirow{5}{*}{\begin{tabular}[c]{@{}c@{}}Performance\\ (Normalized MLU)\end{tabular}}} & Average   & 1.12    & 1.14 & 1.47 \\
\multicolumn{1}{c|}{}                                                                                        & 25th Pct. & 1.07    & 1.09 & 1.18 \\
\multicolumn{1}{c|}{}                                                                                        & 50th Pct. & 1.12    & 1.11 & 1.32 \\
\multicolumn{1}{c|}{}                                                                                        & 75th Pct. & 1.16    & 1.16 & 1.67 \\
\multicolumn{1}{c|}{}                                                                                        & 99th Pct. & 1.28    & 1.33 & 2.46 \\ \hline
\multicolumn{2}{c|}{Memory Usage (GiB)}                                                                                    & \textbf{0.02} & 0.53 & 0.19 \\ \hline
\multicolumn{2}{c|}{\begin{tabular}[c]{@{}c@{}}Time-to-Performance (s)\\ (1.2/1.18/1.16 Normalized MLU)\end{tabular}}             & \textbf{43/53/59} & 82/146/214 & -- \\ \bottomrule
\end{tabular}
}
\caption{Comparing Geminet, HARP, and SaTE on DynGeant.}
\label{tab: comparing harp geminet sate}
\end{table}
\subsection{Comparing in single-cluster topologies}
\label{sec: comparing with other TE schemes}
We compare TE schemes on single-cluster topologies, focusing on performance, the number of neural network parameters, GPU memory usage, and convergence speed.

\begin{figure*}[!ht]
    \subfigure[TE performance on GEANT under original and shuffled paths.]{
    \label{fig: one cluster geant}
    \includegraphics[scale=0.33]{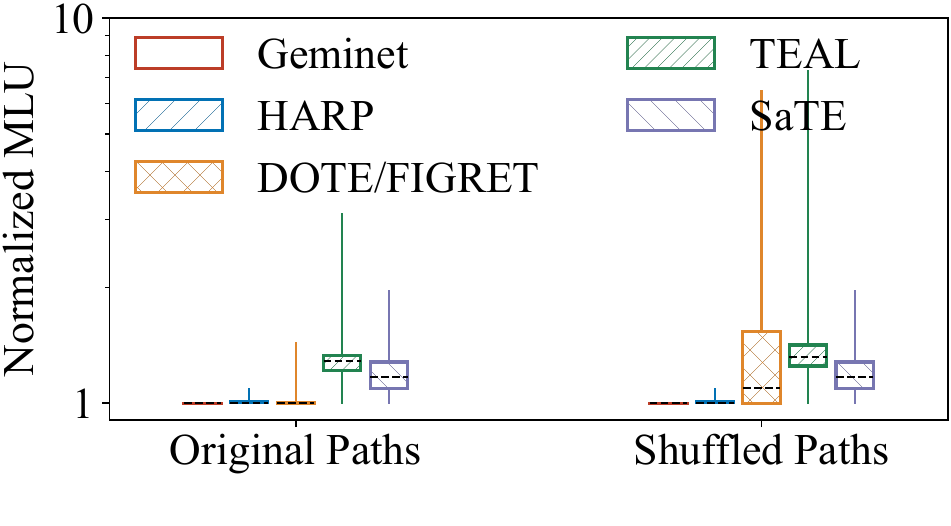}
    }
    \hspace{1mm}
    \subfigure[TE performance on Cogentco under original and shuffled paths.]{
    \label{fig: one cluster cogentco}
    \includegraphics[scale=0.33]{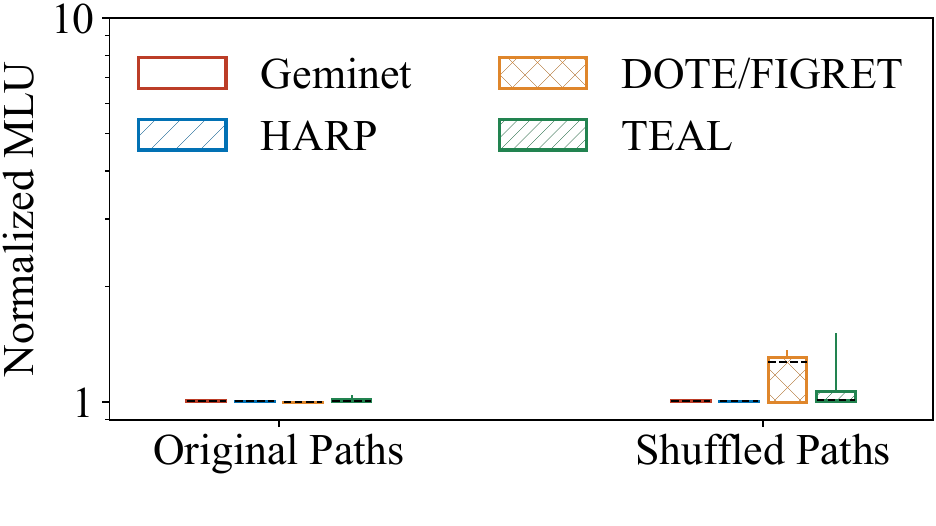}
    }
    \hspace{1mm}
    \subfigure[Coping with different numbers of random link failures on GEANT.]{
    \label{fig: link failure geant}
    \includegraphics[scale=0.33]{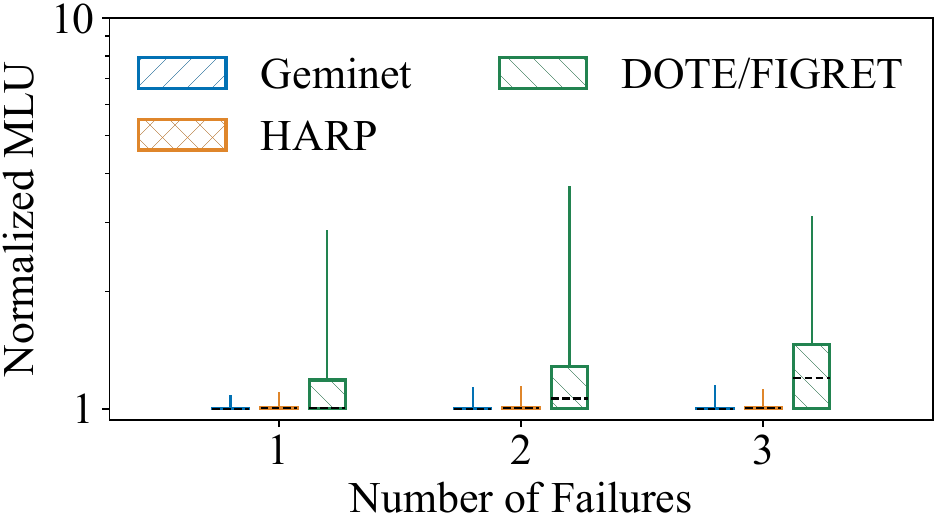}
    }
    \caption{TE performance across different scenarios, including path shuffling (\ref{fig: one cluster geant}, \ref{fig: one cluster cogentco}) and link failures (\ref{fig: link failure geant}). The y-axis shows the MLU normalized by the optimal MLU obtained using Gurobi, and is therefore always greater than 1.}
    \label{fig: performance on shuffled and link failures}
\end{figure*}

\noindent\textbf{Performance.} 
Figure \ref{fig: one cluster geant} and Figure \ref{fig: one cluster cogentco} summarize TE performance for two scenarios: (a) when path orders in training and testing are the same, and (b) when they are shuffled. Geminet and HARP perform well in both cases, while DOTE struggles under shuffled conditions. As discussed in \S \ref{sec: background}, TEAL’s independent optimization for each source-destination pair may sometimes lead to suboptimal results, e.g., as shown in Figure \ref{fig: one cluster geant}. Results for additional topologies are provided in Appendix \ref{sec: additional performance results}, which exhibit similar characteristics.

\begin{figure}[!ht]
    \centering
    \includegraphics[scale=0.17]{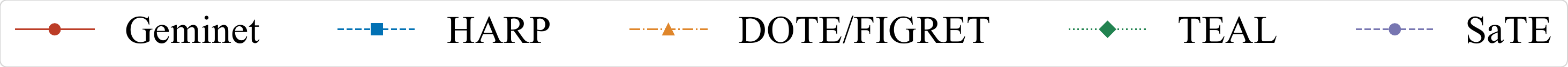}
    
    \subfigure[Comparison of the number of model parameters.]{
    \label{fig: model size}
    \includegraphics[scale=0.193]{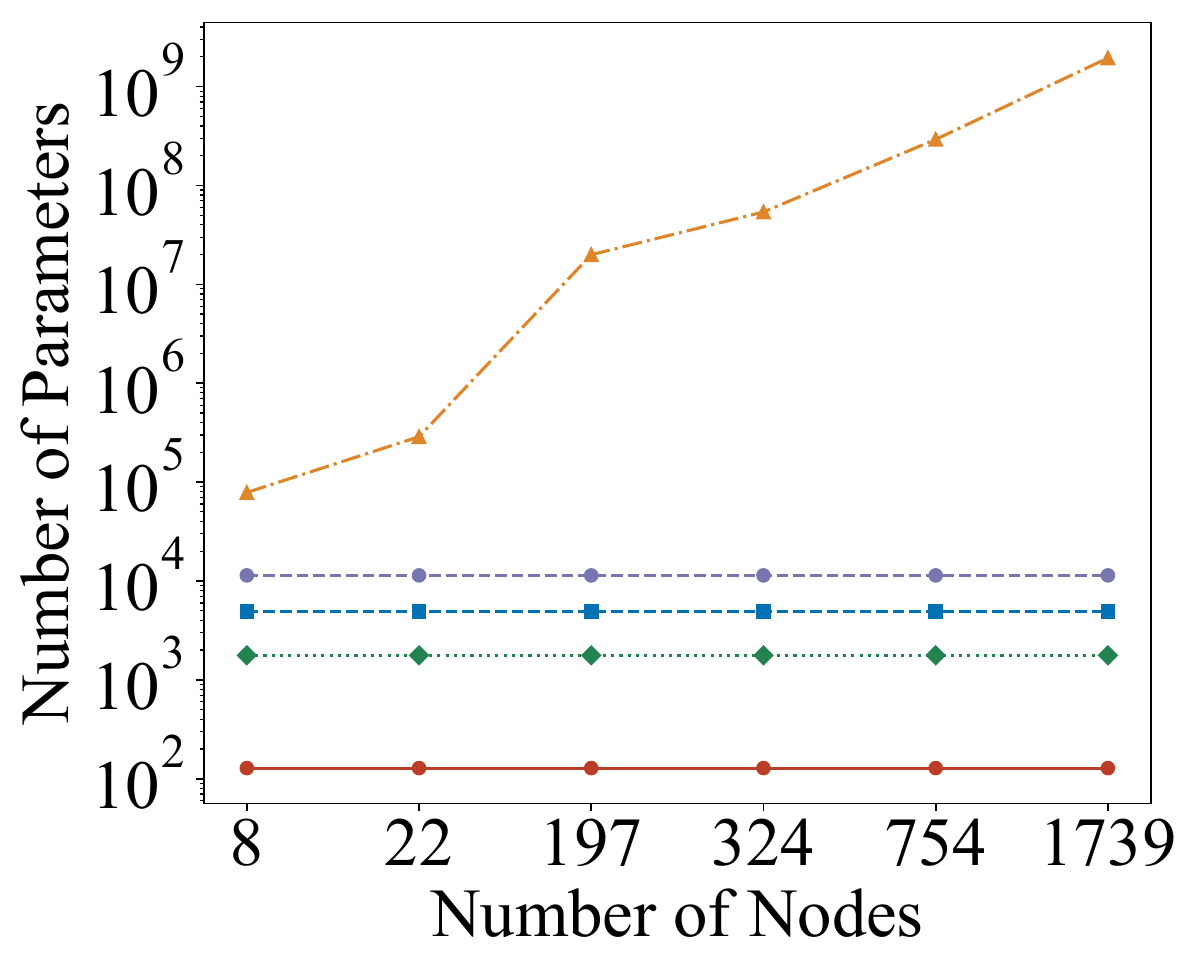}
    }
    \hspace{1mm}
    \centering
    \subfigure[Comparison of Time-to-Performance on KDL.]{
    \label{fig: compare time-to-performance}
    \includegraphics[scale=0.193]{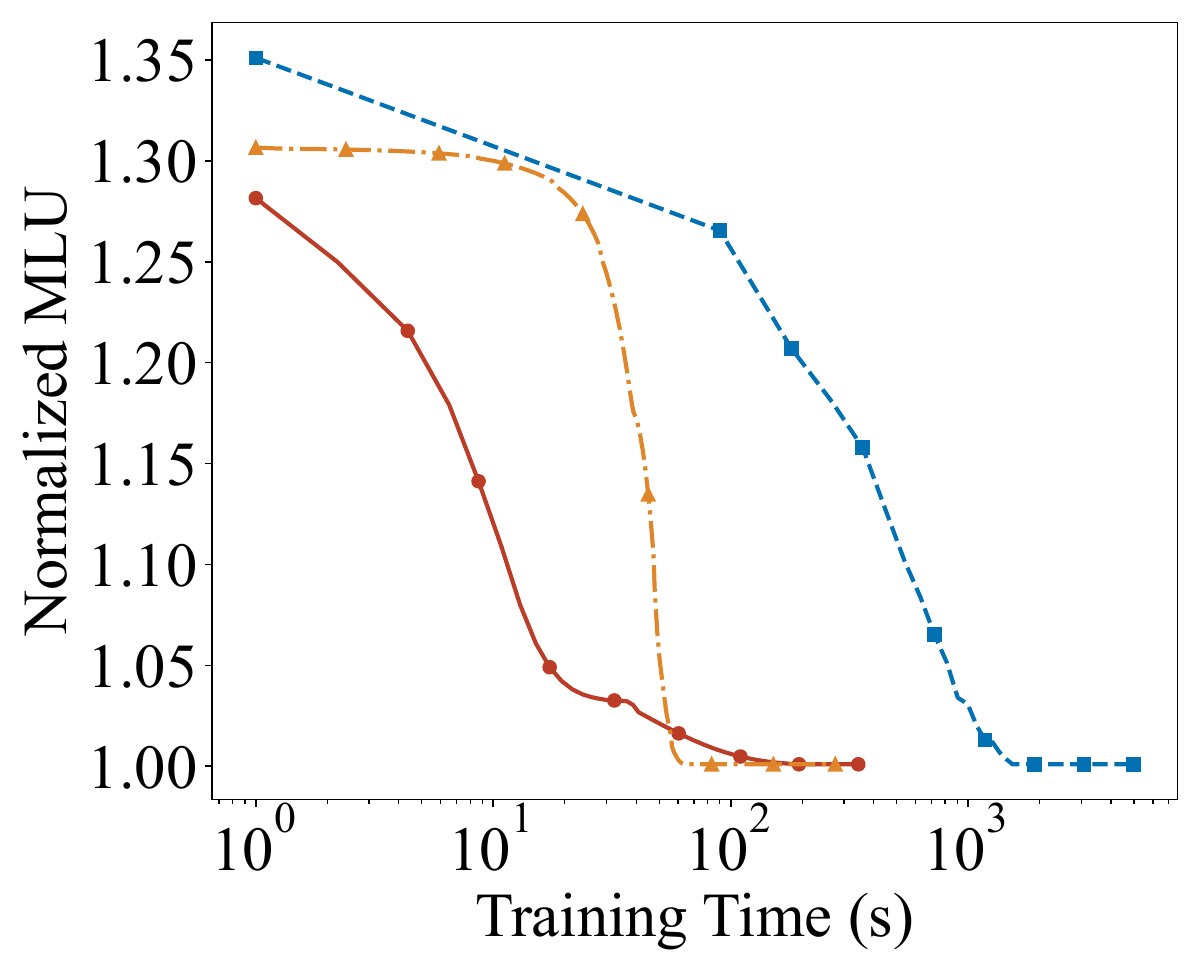}
    }
    \caption{Comparison of the number of model parameters and Time-to-Performance across different TE schemes.}
    \label{fig: compare model size and throughput.}
\end{figure}

\noindent\textbf{Number of neural network model parameters.} In Figure \ref{fig: model size}, we analyze how the number of model parameters for Geminet, HARP, TEAL, SaTE, and DOTE/FIGRET change with topology size. It can be observed that the number of Geminet's model parameters is significantly smaller than the other four methods. For example, on the GEANT topology, the number of parameters in Geminet is only $\mathbf{0.07}$ of TEAL, $\mathbf{0.026}$ of HARP, $\mathbf{0.011}$ of SaTE, and $\mathbf{4\times 10^{-4}}$ of DOTE/FIGRET. Smaller parameter sizes offer many advantages: in addition to requiring less model memory, smaller models typically require less training data, are less prone to overfitting, and may exhibit better generalization performance \cite{tian2022comprehensive}.
The constant parameter counts of TEAL, HARP, and SaTE are expected. These methods represent topology using weight-sharing graph-/set-based modules (e.g., GNN message passing and permutation-invariant attention), whose trainable weights are shared across all nodes/edges/paths; thus, the number of parameters is determined by fixed hyperparameters rather than the topology size.

\noindent\textbf{Memory Usage.} The GPU memory usage results are summarized in Table \ref{tab: memory_usage}. As shown, Geminet has the smallest memory usage among all schemes. Moreover, compared to HARP, another TE scheme capable of handling changing topologies, Geminet significantly reduces memory usage. For instance, on the small-scale GEANT topology, its memory usage is only $25.6\%$ of HARP, and on the large-scale Meta WEB DC ToR topology, its memory usage is just $9.3\%$ of HARP. For HARP, on the KDL topology, it is unable to run even on modern GPUs with 80 GiB of memory. According to our evaluations, even with a batch size of 1 and FP16 training, HARP still consumes 67 GiB of memory on the KDL topology.

To quantify the savings contributed solely by the duality-based edge-level formulation (edge-level vs.\ path-level variables), we conduct an ablation in Appendix~\ref{sec: additional results}: while achieving comparable final TE performance, the edge-level variant reduces training GPU memory by $19.0\%$--$25.6\%$ on large topologies (Cogentco, WEB ToR, and KDL).

\noindent\textbf{Time-to-Performance.} We compare Time-to-Performance of several methods on large-scale KDL topologies, as shown in Figure \ref{fig: compare time-to-performance}. To ensure fairness, we set the batch size of all methods to $1$, since HARP cannot run with larger batches due to out-of-memory issues. As illustrated in Figure \ref{fig: compare time-to-performance}, Geminet converges faster than HARP and performs comparably to DOTE, which only uses a few fully connected layers. Specifically, the times required for Geminet, DOTE, and HARP to reach a normalized MLU of 1.01 are 77s, 56s, and 1357s, respectively. The convergence speed of Geminet is $\mathbf{18\times}$ faster than that of HARP. For a fixed topology with sufficient memory and training data, DOTE can converge fast, since it only has a few fully connected layers, and its forward and backward passes are fast due to the shallow depth. We omit TEAL and SaTE from Figure~\ref{fig: compare time-to-performance} for readability, since RL-based methods typically take much longer to converge than the baselines shown. For example, TEAL requires about $8$ hours to converge on KDL (consistent with TEAL’s report).
\subsection{Handling link failure}
\label{sec: handling link failure}
In this section, we compare the ability of ML-based TE schemes to handle link failures. Since TEAL sometimes performs sub-optimally, we focus only on comparing Geminet, HARP, and DOTE. For Geminet and HARP, handling link failures can be regarded as a change in edge capacity. DOTE’s strategy for handling link failures is local rescaling \cite{perry2023dote}. Specifically, traffic on unavailable tunnels is rerouted to the surviving tunnels in proportion to the split ratios.

Figure \ref{fig: link failure geant} shows the impact of randomly selected link failures in the GEANT. Geminet performs on par with HARP in handling link failures. Moreover, ML-based TE schemes like Geminet and HARP, which support variable topologies, outperform DOTE, which relies on rerouting. Results for other topologies in Appendix \ref{sec: additional link failure results} exhibit similar trends.

\begin{figure*}
\hspace{2mm}
\begin{minipage}{0.32\textwidth}
    \centering
    \includegraphics[scale=0.3]{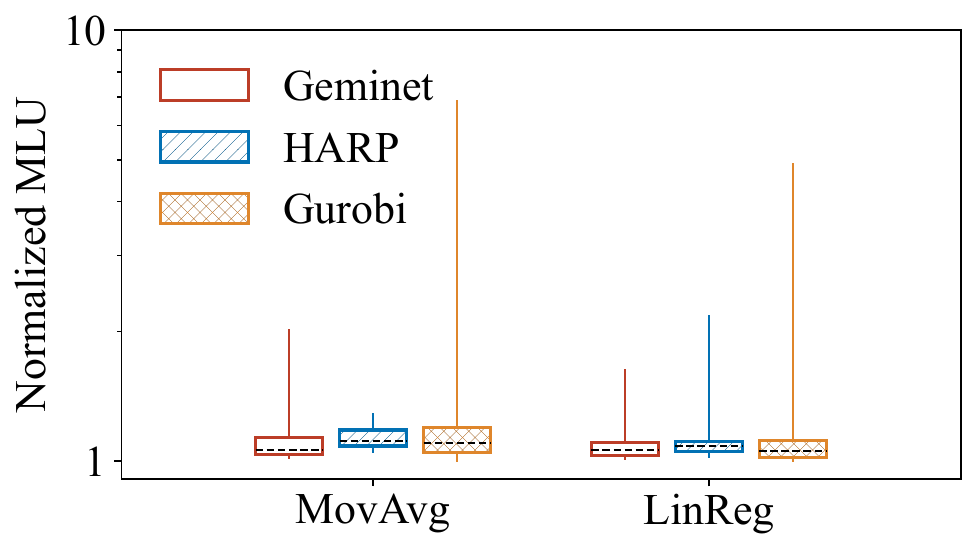}
    \caption{Comparison of Geminet, HARP, and Gurobi performance on DynGeant with predicted traffic demands.}
    \label{fig: dygeant_pred}
\end{minipage}
\hspace{2mm}
\begin{minipage}{0.32\textwidth}
    \centering
    \includegraphics[scale=0.3]{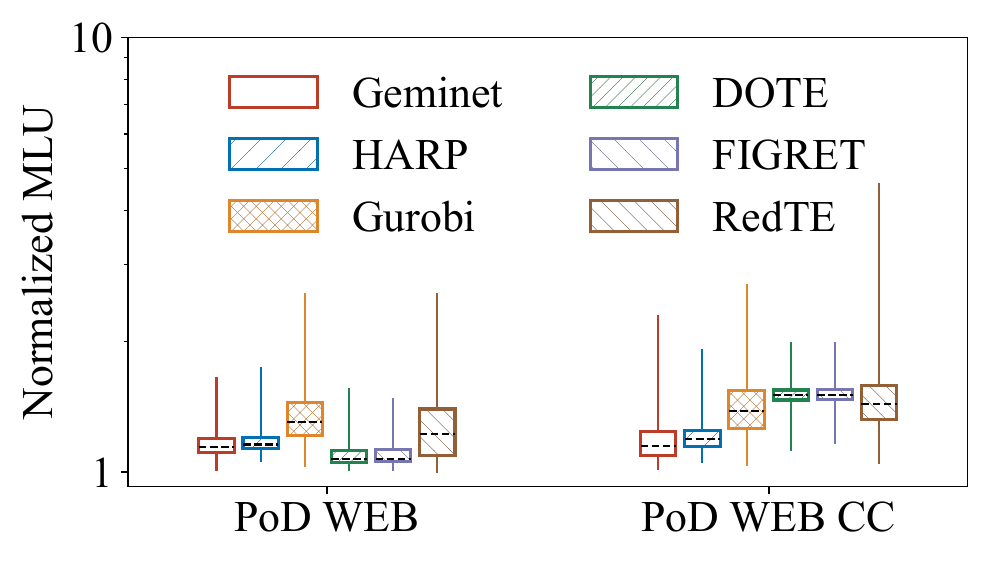}
    \caption{Comparison of schemes on PoD WEB under traffic uncertainty, where CC represents changing edge capacity.}
    \label{fig: compare end-to-end in web}
\end{minipage}
\begin{minipage}{0.32\textwidth}
    \centering
    \includegraphics[scale=0.3]{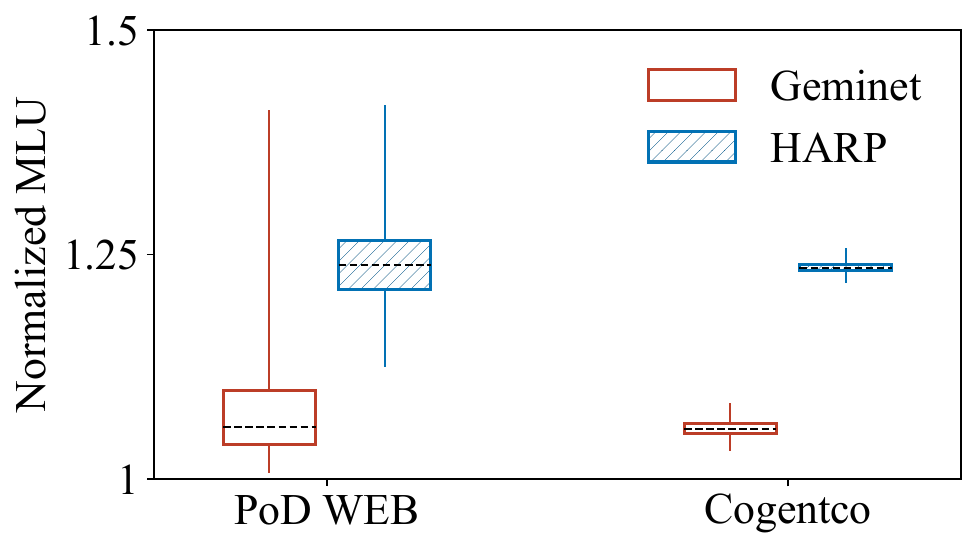}
    \caption{Cross-topology generalization results with models trained on GEANT and tested on different topologies.}
    \label{fig: pod_cogentco_comparison}
\end{minipage}
\end{figure*}

\subsection{Handling traffic uncertainty}
\label{sec: geminet under traffic uncertainty}
\textbf{Performance comparison across TE schemes.} Our results so far have evaluated schemes using exact traffic matrices. However, in practice, both optimization solvers and ML-based schemes operate with predicted traffic matrices. Consequently, we next aim to evaluate the performance of our algorithm under traffic uncertainty.
Similar to HARP \cite{alqiam2024transferable}, we first predict the traffic matrices and then feed these predicted matrices to Geminet. During the training phase, TE configurations are generated using the predicted matrix, but the loss is computed using the actual matrix. In the inference phase, only the predicted matrix is available to Geminet. For traffic prediction, we employ two methods, as described in \cite{alqiam2024transferable}: (i) MovAvg: The predicted value is the average of the corresponding historical window. (ii) LinReg: The predicted value is based on linear regression applied to the historical window. The historical window length is set to $12$ \cite{alqiam2024transferable,perry2023dote}.
The performance comparison of Geminet, HARP, and Gurobi on the dynamic topology DynGeant under predicted traffic demands is summarized in Figure \ref{fig: dygeant_pred}. First, similar to findings in \cite{alqiam2024transferable}, ML-based TE demonstrates an ability to learn robustness against prediction errors compared to Gurobi. Second, Geminet shows a lower mean and 50th percentile compared to HARP. However, in extreme cases, such as the top end of the whiskers in the MovAvg boxplot, Geminet exhibits higher values. This is due to their differing adjustment strategies: Geminet adjusts at the edge level, while HARP operates at the path level. Edge-level adjustments aggregate flows from multiple paths, allowing prediction errors to partially cancel out, which generally benefits Geminet. However, when prediction errors across paths align, this cancellation effect diminishes, leading to greater errors and causing Geminet to underperform in extreme cases.

To compare Geminet and HARP with end-to-end approaches like DOTE and FIGRET, which directly map historical traffic matrices to next-time TE decisions, we evaluate them on the WEB DC topology under two scenarios: (i) fixed edge capacities and (ii) changing edge capacities, with FIGRET’s traffic burst handling weight set to $0.1$. The results, summarized in Figure \ref{fig: compare end-to-end in web}, show that end-to-end methods outperform Geminet and HARP in handling traffic uncertainty when edge capacities are fixed. However, their performance degrades significantly when edge capacities change (PoD WEB CC), as they assume a static topology. Additional results on other topologies are provided in Appendix~\ref{sec: additional traffic uncertainty results}.

\noindent\textbf{Discussion on RedTE performance.}
We observe that RedTE performs worse than expected in Figure~\ref{fig: compare end-to-end in web} even when there is no topology change. First, we clarify that RedTE follows the standard one-step-ahead control setting: it observes the state at time $t$ and outputs an action applied to the demand at time $t{+}1$, rather than applying actions to the same instantaneous demand (consistent with the RedTE paper and official code \cite{RedTE}). This formulation is more effective when the control interval is short; at the sampling granularity of the public traces we use, the demand can change non-trivially between consecutive steps. Second, we reproduce RedTE using the authors’ open-source implementation and recommended hyperparameters; however, RL-based methods are harder to train and tune and can be sensitive to training stability, so our results may underestimate RedTE’s best achievable performance, which we acknowledge as a limitation. Additionally, we provide an analysis of RedTE's large memory footprint in Appendix~\ref{sec: redte_memory}.

\subsection{Computation Time}
\label{sec: computation time}
We analyze the computation time of ML-based TE schemes and Gurobi. We create the model and run inference directly for ML-based TE schemes with too large memory requirements to train, as whether the model has been trained does not affect inference time. The results are summarized in Figure \ref{fig: comparison of computation times}. Figure~\ref{fig: comparison of computation times} includes both WAN and DC topologies; the 8-node and 324-node cases correspond to DC networks. We measure the end-to-end latency from receiving the topology, traffic, and paths to outputting splitting ratios, including preprocessing overhead (e.g., bipartite map construction in SaTE to form the input of its neural network).
From the results, we draw the following findings: (i) ML-based TE schemes have an advantage over Gurobi in computation time, being faster than two orders of magnitude for large-scale topologies such as KDL. (ii) The two TE schemes, Geminet and HARP, which are capable of handling changing topologies, require iterative adjustments, making them slower than TEAL and DOTE. (iii) Among the ML-based TE schemes designed for changing topologies, Geminet is faster than HARP and SaTE. On the KDL topology, Geminet achieves a computation speed $2.3\times$ that of HARP. (iv) For ASN, its average path length is shorter than KDL, resulting in a sparser path-to-edge mapping. Thus, the computation time does not increase significantly for TEAL, Geminet, and HARP, which all process path-edge relationships during inference. In contrast, computation time increases on ASN for DOTE, whose model size depends only on the number of nodes. A similar effect holds for the DC topologies, where paths are typically shorter and the path--link mapping is sparser.
In addition, to better support deployment considerations, we report Geminet's peak GPU memory footprint during inference in Appendix~\ref{sec: inference_memory}.
\subsection{Handling drastically different topology}
\label{sec: handling drastically different topology}
In this section, we evaluate the performance of Geminet and HARP when trained on one topology and tested on a completely different one. SaTE is omitted here due to its suboptimal performance. Specifically, both models are trained on GEANT and tested on Meta WEB PoD-level and Cogentco topologies, with results summarized in Figure \ref{fig: pod_cogentco_comparison}. We observe that when the test topology differs substantially from the training topology, both Geminet and HARP exhibit noticeable performance degradation, which is a typical manifestation of domain shift in machine learning. Consequently, training on a small topology and directly deploying on a large and structurally different topology without adaptation is unlikely to yield optimal TE performance. To rule out that the above trend is an artifact of a particular train/test direction, Appendix~\ref{sec: reverse cross-topology} (Figure~\ref{fig: reverse_pod_cogentco_comparison}) reports the reverse transfer setting (trained on Cogentco and tested on Meta WEB PoD-level and GEANT), which shows a similar degradation. Notably, we do not observe a consistent winner between Geminet and HARP across transfer directions; the relative performance depends on the specific source--target pair and the induced domain shift.
\subsection{Visualization of the Update Process}
\label{sec: Visualization of the Update Process}
We provide several examples to visualize the update process. Results on Meta WEB PoD-level and GEANT are shown in Figure~\ref{fig: update curve}. For clarity, we select the 1st, 100th, and 200th test samples as illustrative cases. As can be seen, the update operator iteratively drives the solution towards the optimum within only a few steps. This not only offers an intuitive visualization but also empirically verifies that the learned operator effectively captures the desired update dynamics.

\noindent\textbf{Convergence behavior.}
Figure~\ref{fig: update curve} empirically demonstrates a convergent optimization trajectory: the normalized MLU decreases over iterations and gradually plateaus, indicating diminishing improvements after several update steps. Due to the black-box nature of the learned update operator, providing rigorous theoretical convergence/stability bounds is non-trivial. Therefore, we unroll a fixed number of iterations $T$ and rely on empirical evidence to validate stability.


\noindent\textbf{Controlled probe of the learned updater.}
To better understand what the core network learns, we conduct a controlled input sweep on MLP 2: we fix link capacity and jointly vary the $\mathrm{LU}/\mathrm{MLU}\in(0,1]$ and the current dual variable $\lambda$, with both ranges calibrated from real inference trajectories (Appendix Figure \ref{fig: edge_mlp_probe}). The resulting response surface shows that the network tends to increase the dual variable of edges whose utilization approaches the bottleneck, thereby raising their routing costs and steering traffic away from congested links.
\section{Discussion}
\label{sec: discussion}
\noindent\textbf{Extensibility to other objectives.} Other optimization objectives can generally be categorized into two types. The first includes convex objectives, such as maximizing total throughput, which can often be formulated similarly to MLU. Geminet can be adapted to optimize these objectives with only minor modifications. The second category consists of non-convex and non-differentiable objectives, such as minimizing flow completion time or packet loss rate, which often lack explicit analytical formulations. These present a greater challenge for existing ML-based approaches, which are currently not well-suited to handle such objectives. A potential solution is to integrate a simulator that can approximate these objectives and provide feedback, allowing an ML-based method to refine its strategy. We acknowledge that when the objective does not admit a closed-form analytical expression, Geminet cannot be directly applied, and additional neural modules may be necessary. We leave this exploration for future work.

\noindent\textbf{Improving Geminet’s solution time.} As discussed in \S\ref{sec: computation time}, Geminet takes longer to compute solutions than DOTE, primarily because DOTE produces results directly through fully connected layers, whereas Geminet relies on an adjustment process. This process requires computing gradient-related quantities, such as MLU, which involves sparse matrix multiplications. These computations bottleneck Geminet’s solution time. This is the cost of supporting changing topologies, which remains an area for future optimization; in particular, standard accelerations for sparse matrix multiplications (e.g., mixed precision and more efficient compressed sparse formats) may help reduce this overhead.

\noindent\textbf{End-to-end training of Geminet with prediction tasks.}
Figure~\ref{fig: compare end-to-end in web} (\S\ref{sec: geminet under traffic uncertainty}) shows that in static topologies, end-to-end methods (DOTE/FIGRET), which directly map historical traffic matrices to TE decisions via joint training, outperform two-stage pipelines (Geminet/HARP) under traffic uncertainty. This benefit largely comes from mitigating the mismatch between prediction losses (e.g., MSE) and TE objectives (e.g., MLU) \cite{perry2023dote}. Extending this idea to TE schemes that handle changing topologies is non-trivial, since their iterative procedures require accurate demands to evaluate MLU; enabling end-to-end training in this setting remains open.

\noindent\textbf{How to handle drastic topology changes.} As shown in \S\ref{sec: handling drastically different topology}, Geminet’s performance declines when the test topology differs markedly from the training one. A practical remedy is to apply fine-tuning when adapting to drastic topology changes. We acknowledge that we do not yet have a precise quantitative definition of what constitutes a \emph{drastic} shift; qualitatively, we consider a shift to be drastic when it changes the structural characteristics of the topology (e.g., connectivity patterns or overall sparsity), rather than incremental updates such as adding/removing a few nodes or links or adjusting link capacities. In practice, engineers may adopt periodic fine-tuning when such structural changes are detected. Thanks to Geminet’s small model size, we expect only a few iterations would suffice for convergence, and we leave this for future work.

\section{Related work}
TE is widely used in WANs and data centers. However, as network topologies expand, TE optimization becomes increasingly time-consuming. To address this, efforts have focused on accelerating TE, including scaling optimization via decomposition methods \cite{abuzaid2021contracting,ghosh2013scalable,narayanan2021solving,jiang2022flexile}. However, these methods are constrained by limited subproblem partitioning, balancing runtime and TE performance, but achieving only modest parallelism and speedup. These limitations have spurred solver-free accelerators, which optimize per-SD split ratios via balanced binary search \cite{mao2025atro,mao2025fast}, as well as ML-based TE.

ML-based TE has been extensively studied \cite{liu2020automated,valadarsky2017learning,xu2023teal,perry2023dote,liu2024figret,alqiam2024transferable,xu2018experience,mohammed2019deeproute,gui2024redte}. TEAL \cite{xu2023teal} leverages ML to accelerate TE decision-making, while DOTE \cite{perry2023dote} optimizes routing under traffic uncertainty using historical traffic demand. FIGRET \cite{liu2024figret} enhances robustness against traffic bursts by incorporating an additional loss function. Then, HARP \cite{alqiam2024transferable} introduces ML-based TE for topology variations, but its high computational cost limits practicality. In this work, we propose a lightweight ML-based TE solution for dynamic topologies, making it more suitable for real-world deployment.

For link failures, TE schemes are either proactive or reactive. Proactive methods guarantee congestion-free operation across all predefined failure scenarios \cite{liu2014traffic,jiang2020pcf,wang2010r3} or enough scenarios to meet a percentile target \cite{bogle2019teavar,chang2019lancet}, often via conservative bandwidth allocation. Reactive methods adapt after failures via local rescaling or explicit recomputation \cite{abuzaid2021contracting,kumar2018semi}. These strategies can be combined, and recent ML-based TE further incorporates fault tolerance via failure-aware losses \cite{liu2025faute}. Our approach is reactive and adapts to post-failure topologies.


\section{Conclusion}
We introduce Geminet, a framework that leverages a duality-based update process for lightweight traffic engineering in dynamic topologies. Geminet decouples the neural network from topology by learning a topology-agnostic update operator and applies duality theory to transform the optimization problem from the path level to the edge level. Our evaluation demonstrates that Geminet substantially reduces computational and memory overhead while maintaining performance, showing its potential for deployment in production.

\section*{Acknowledgments}
We sincerely thank our shepherd, T. S. Eugene Ng, and the anonymous reviewers for their constructive feedback. This work was supported by the NSF China (No. U25B2039) and the Zhongguancun Academy (No. C20250205).

\bibliographystyle{plain}
\bibliography{reference}
\appendix

\section*{Appendix}
\section{Notation table}
\label{sec: notation table}
We tabulate the notations in Table \ref{tab: notation table}.
\begin{table}[h]
\scalebox{0.83}{
\begin{tabular}{|l|l|}
\hline
\textbf{Notation}          & \textbf{Description}                                                                                                                                                                                       \\ \hline
$G(V,E,c)$                 & \begin{tabular}[c]{@{}l@{}}Network topology, $V$ is node set, $E$ is edge \\ set, and c assigns capacities to edges\end{tabular}                                                                           \\ \hline
$D$                        & \begin{tabular}[c]{@{}l@{}}Demand matrix, where $D_{ij}$ denotes the\\ traffic from $i$ to $j$\end{tabular}                                                                                                \\ \hline
$P$                        & \begin{tabular}[c]{@{}l@{}}Network paths, where $P_{st}$ denotes the set \\ of network paths through which the source \\ $s$ communicates with destination $t$\end{tabular}                                \\ \hline
$r_p$                      & \begin{tabular}[c]{@{}l@{}}The split ratio of the traffic demand from \\ $s$ to $t$ forwarded along path $p$\end{tabular}                                                                                  \\ \hline
$\mathcal{R}$              & TE configuration                                                                                                                                                                                           \\ \hline
$m$                        & Max Link Utilization                                                                                                                                                                                       \\ \hline
$U(e)$                     & Link utilization of edge $e$                                                                                                                                                                               \\ \hline
$\lambda_e$                & \begin{tabular}[c]{@{}l@{}}Lagrange dual variable associated with \\ edge $e$\end{tabular}                                                                                                                 \\ \hline
$L(\mathcal{R},m,\lambda)$ & Lagrangian function                                                                                                                                                                                        \\ \hline
$\Phi$                     & The set of all paths                                                                                                                                                                                       \\ \hline
$\xi^{(|\Phi|,|E|)}$       & \begin{tabular}[c]{@{}l@{}}Path-to-Edge Matrix: A binary matrix \\ indicating whether a given path $i$ \\ contains edge $j$. Specifically,  $\xi_{ij} = 1$ \\ if edge $j$ is part of path $i$\end{tabular} \\ \hline
$\Psi^{(|\Phi|,1)}$        & \begin{tabular}[c]{@{}l@{}}A vector representing the potential load \\ associated with each path\end{tabular}                                                                                              \\ \hline
$\Omega$                   & The set of all source-destination pairs                                                                                                                                                                    \\ \hline
$S$                        & \begin{tabular}[c]{@{}l@{}}The number of sub-traffic matrices obtained \\ by partitioning the traffic matrix.\end{tabular}                                                                                 \\ \hline
\end{tabular}
}
\caption{Notations used in this paper.}
\label{tab: notation table}
\end{table}
\section{Mathematical Preliminaries}
\label{sec: mathematical preliminaries}
\subsection{Summation order exchange detail}
\label{sec: summation order exchange detail}
We explain the transition from the second to the third line in Equation (\ref{equ: lagrangian partial}). We begin by writing out the second term of the second line in Equation (\ref{equ: lagrangian partial}), which is given by:
\begin{equation}
\sum_{e\in E}\sum_{s,t\in V}\sum_{p\in P_{st}, p \owns e}\lambda_e D_{st}r_p.
\label{equ: second term of lagrangian line 2}
\end{equation}
To facilitate this transition, we introduce the indicator function $I(p,e)$, which equals 1 if path p includes edge e (denoted \(p \owns e\)), and 0 otherwise. Incorporating $I(p,e)$ into the Equation (\ref{equ: second term of lagrangian line 2}), we rewrite it as:
\begin{equation}
\sum_{e\in E}\sum_{s,t \in V}\sum_{p\in P_{st}}I(p,e)\lambda_e D_{st}r_p.
\label{equ: line 2 with indicator}
\end{equation}
Since the summations in Equation (\ref{equ: line 2 with indicator}) are independent and do not interfere with each other, we can rearrange the order of summation as follows:
\begin{equation}
    \sum_{s,t \in V}\sum_{p \in P_{st}}\sum_{e\in E}I(p,e)\lambda_e D_{st}r_p.
\label{equ: line 2 with indicator rearrange}
\end{equation}
Given that the indicator function $I(p,e)$ equals $1$ only when $p \owns e$, the summation $\sum_{e\in E}I(p,e)$ simplifies to $\sum_{e \in p}$. Consequently, Equation (\ref{equ: line 2 with indicator rearrange}) can be rewritten as:
\begin{equation}
    \sum_{s,t \in V}\sum_{p \in P_{st}}\sum_{e \in p} \lambda_e D_{st} r_p,
\end{equation}
which is the second term of the third line in Equation (\ref{equ: lagrangian partial}).

\subsection{Primal optimal from Lagrangian}
\label{sec: primal optimal from lagrangian}
\begin{lemma}
    The solution of the Equation (\ref{equ: te lagrangian with optimal dual}) is primal optimal.
\label{lemma: primal optimal from lagrangian optimal}
\end{lemma}
To prove LEMMA \ref{lemma: primal optimal from lagrangian optimal}, we begin by presenting two foundational lemmas from duality theory \cite{boyd2004convex}:
\begin{lemma}
    For linear programming problems, strong duality always holds if the primal problem is feasible and bounded \cite{boyd2004convex}.
\label{lemma: strong duality}
\end{lemma}
\begin{lemma}
\label{lemma: uniqueness and optimality}
    If strong duality holds and the optimal dual variables $\lambda^*,\nu^*$ are known, and the minimizer of $L(x,\lambda^*,\nu^*)$, i.e., the solution of 
    \begin{equation}
        \text{minimize}\quad f_0(x) +\sum_{i=1}^m \lambda_i^* f_i(x) + \sum_{i=1}^p \nu_i^* h_i(x),
        \label{equ: minimize lagrangian with optimal dual}
    \end{equation}
    is unique. Then if the solution of Equation \ref{equ: minimize lagrangian with optimal dual} is primal feasible, it must be primal optimal \cite{boyd2004convex}.
\end{lemma}

Given LEMMA \ref{lemma: strong duality} and LEMMA \ref{lemma: uniqueness and optimality}, we proof LEMMA \ref{lemma: primal optimal from lagrangian optimal} as follow:
\begin{proof}
    Given that the original problem specified by Equation (\ref{equ: te opt formulation}) is a linear programming problem, we apply LEMMA \ref{lemma: strong duality} to assert that strong duality holds. Furthermore, the minimization of the Lagrangian, as formulated in Equation (\ref{equ: te lagrangian with optimal dual}), constitutes another linear problem. Since linear problems are a subset of convex optimization problems, the solution is unique. Therefore, according to LEMMA \ref{lemma: uniqueness and optimality}, solving Equation (\ref{equ: te lagrangian with optimal dual}) yields the optimal solution to the original problem.
\end{proof}

\subsection{Dual problem of TE optimization}
\label{sec: dual problem of TE optimization}
In this section, we present the dual problem of the primal problem defined by Equation (\ref{equ: te opt formulation}), The dual function $g(\lambda)$ is defined as the infimum of the Lagrangian $L(\mathcal{R}, m, \lambda)$ with respect to the primal decision variables $\mathcal{R}$ and $m$: $g(\lambda) = \inf_{\mathcal{R}, m} L(\mathcal{R}, m, \lambda)$.

\noindent(i) Minimization with Respect to $m$:
The Lagrangian contains the following term involving $m$:
\[
\bigg( 1 - \sum_{e \in E} \lambda_e c(e) \bigg) m.
\]
To ensure the dual function $g(\lambda)$ remains bounded, we analyze the behavior of this term in the following cases:
\begin{itemize}[leftmargin=*,topsep=0pt, partopsep=0pt]
    \item If $1 - \sum_{e \in E} \lambda_e c(e) > 0$, the term $\big(1 - \sum_{e \in E} \lambda_e c(e)\big) m$ becomes unbounded as $m \to -\infty$, causing $L(R, m, \lambda)$ to diverge to $-\infty$. This makes the dual problem infeasible.
    \item  If $1 - \sum_{e \in E} \lambda_e c(e) < 0$, the term becomes unbounded as $m \to +\infty$, leading to a similar divergence of $L(R, m, \lambda)$ to $-\infty$.
    \item  Therefore, the only feasible case is when: $1 - \sum_{e \in E} \lambda_e c(e) = 0$
\end{itemize}

\noindent (ii) Minimization with Respect to $\mathcal{R}$: The part of the Lagrangian involving $\mathcal{R}$ is given by:
\[
\sum_{s,t\in V}\sum_{p\in P_{st}}\sum_{e\in p} \lambda_e D_{st}r_p.
\]
Minimizing this term with respect to R must satisfy the primal constraints $r_p \geq 0$ and $\sum_{p \in P_{st}} r_p = 1$. As analyzed in \S \ref{sec: harness duality in TE}, this is minimized when all the weight is placed on the path $p \in P_{st}$ with the smallest $\sum_{e \in p} \lambda_e$.

\noindent (iii) In summary, the dual problem is shown in Equation (\ref{equ: dual problem of te}).

\begin{equation}
\boxed{
\begin{aligned}
    &\mathop{\text{maximize}}_{\lambda}\quad&&\sum_{s,t\in V} D_{st}\bigg(\min_{p\in P_{st}}(\sum_{e\in p}\lambda_e)\bigg)\\
    &\text{subject to}\quad&&1-\sum_{e\in E} \lambda_e c(e) = 0,\\
    & &&\lambda_e \geq 0,\forall e\in E.
\end{aligned}
}
\label{equ: dual problem of te}
\end{equation}



\section{Additional Details}
\subsection{Details of self-attention}
\label{sec: self-attention}
In this section, we briefly introduce the computation process of self-attention and explain why using transformers to obtain path embeddings generates intermediate variables whose size is proportional to the square of the path length.

The self-attention mechanism processes an input sequence $X \in \mathbb{R}^{l \times d}$, where $l$ is the sequence length and $d$ is the feature dimension, to compute a weighted output that captures contextualized representations of each position in the sequence. The process involves several steps:

\noindent\textbf{1. Projection into Query, Key, and Value:} The input $X$ is transformed into three learned representations: Query ($Q$), Key ($K$), and Value ($V$). These are computed as: $Q=XW_Q, K=XW_K, V=XW_V$, where $W_Q,W_K,W_V\in \mathbb{R}^{d\times d_k}$ are learnable projection matrices, and $Q,K,V\in\mathbb{R}^{l\times d_k}$.

\noindent\textbf{2. Computing the Attention Matrix:} To model the relationships between all sequence positions, pairwise similarity scores between $Q$ and $K$ are calculated using a scaled dot-product:
\begin{equation}
    A = \mathrm{Softmax}(\frac{QK^\mathrm{T}}{\sqrt{d_k}}),
\end{equation}
where $A\in \mathbb{R}^{l\times l}$ represents the attention weights. Each entry in $A$ reflects the importance of one sequence position to another, normalized using the softmax function. 

\noindent\textbf{3. Weighted Summation with the Value Matrix} The attention matrix $A$ is then multiplied with the Value matrix $V$ to compute the final output.

The intermediate variable that grows with the square of the Path length is generated during the computation of the attention matrix, given that the sequence length in HARP is equal to the Path length.
\subsection{Pseudocode for adjusting dual variable}
\label{sec: Pseudocode for adjust dual variable}
\begin{algorithm}[ht]
\SetAlgoLined
\KwIn{Path-to-Edge matrix $\xi^{(|\Phi|, |E|)}$, Capacity vector $C^{(|E|,1)}$, Path traffic demand $DM^{(|\Phi|,1)}$, Initial dual variable $\lambda^{(|E|,1)}$, Number of adjustment iteration $L$, Number of source-destination Paris $|\Omega|$, Number of paths per pair $K$.}
\KwOut{Adjusted dual variable $\lambda^{(|E|,1)}$.}
\BlankLine
\caption{Adjust dual variable $\lambda$}
\label{algorithm: Pseudocode for adjust dual variable}
Let $\mathcal{N}$ be a neural network with input dimension 4 and output dimension 1.\\
\For{i in $0,1,\dots,L$}
{
    $\text{Sum lambda on path } \Lambda^{(|\Phi|,1)} = \xi^{(|\Phi|,|E|)} \times \lambda^{(|E|,1)}$;\\
    \tcp{$\Lambda_{\text{reshape}}^{(|\Omega|,K)}$ is the reshaped form of $\Lambda^{(|\Phi|,1)}$. The Softmin operation is applied along the last dimension. For a detailed explanation of the Softmin operation, please refer to Appendix \ref{sec: Pseudocode for adjust dual variable}.}
    $\text{TE configuration }\mathcal{R}_{\text{reshape}}^{(|\Omega|,K)}=\text{Softmin}(\Lambda_{\text{reshape}}^{(|\Omega|,K)})$; \label{alg line: softmin}\\ 
    $\text{Flow on edge }\mathcal{E}^{(|E|,1)}= \xi^\text{T}\times(\text{DM}^{(|\Phi|,1)}\odot\mathcal{R}^{(|\Phi|,1)})$;\\
    $\text{MLU }m = \max\left(\mathcal{E}^{(|E|,1)}/C^{(|E|,1)}\right)$;\\
    \tcp{$m_{\text{repeat}}$ is the repetition of $m$ for $|E|$ times. The \text{Concatenate} operation is performed along the last dimension.}
    $\text{Neural Network Input $\mathcal{I}^{(|E|,4)}$}= \text{Concatenate}(\mathcal{E}^{(|E|,1)},C^{(|E|,1)},\lambda^{(|E|,1)},m_{\text{repeat}}^{(|E|,1)})$;\\
    $\Delta\lambda^{(|E|,1)}=\mathcal{N}(\mathcal{I}^{(|E|,4)})$;\\
    $\lambda^{(|E|,1)}=\lambda^{(|E|,1)}+\Delta\lambda^{(|E|,1)}$;
}
\Return $\lambda^{(|E|,1)}$
\end{algorithm}
We demonstrate how to adjust the dual variable $\lambda$ in Pseudocode \ref{algorithm: Pseudocode for adjust dual variable}. Our input also includes an additional term, $\lambda_e$, as it helps the neural network determine the magnitude of $\Delta \lambda_e$ based on the value of $\lambda_e$, which is equivalent to selecting an appropriate step size in the gradient descent process. The details of Softmin Operation in Line \ref{alg line: softmin} are as follows:

Softmin is a commonly used mathematical operation typically applied to tasks such as normalization or weight assignment in deep learning. The Softmin operation assigns larger weights to smaller input values and smaller weights to larger input values.\\
Formula: Given an input vector $x=[x_1,x_2,\dots,x_n]$, the output of the Softmin operation is computed as:
\begin{equation}
    y_i = \frac{e^{-x_i}}{\sum_{j=1}^n e^{-x_j}}.
\end{equation}
Characteristics:
\begin{itemize}[leftmargin=*]
    \item \textbf{Normalized output:} Softmin operation ensures that the output is always greater than 0 and the sum of the outputs equals 1. This satisfies the constraints in Equation (\ref{equ: formulate for each pair}).
    \item \textbf{Smaller input values correspond to larger output weights:} Due to the negative exponent, Softmin maps smaller input values to larger weights. This aligns with the requirements of the objective function in Equation (\ref{equ: formulate for each pair}).
\end{itemize}
\subsection{Training on sub-traffic matrices (Optional)}
\label{sec: training on sub-pairs}
While Geminet is already lightweight—requiring significantly less memory than prior schemes—we additionally introduce an \textbf{\textit{optional}} strategy to reduce training overhead further. This strategy is not essential to the core design of Geminet but can be applied when memory constraints are particularly stringent. In the evaluation section, results are reported without this strategy unless explicitly stated to highlight its effect.
\begin{figure}[!ht]
    \centering
    \includegraphics[scale=0.7]{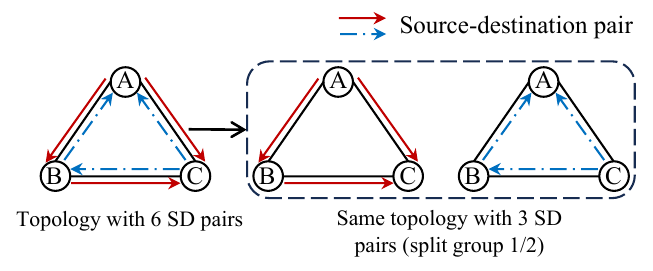}
    \caption{Traffic matrix partitioning ($S=2$).}
    \label{fig: split_pair}
\end{figure}

\noindent\textbf{How to partition the traffic matrix.} Since we learn the update operator, which is independent of the structure of the traffic matrix, we can partition the traffic matrix into sub-traffic matrices to further reduce memory consumption. This involves splitting the source-destination (SD) pairs into multiple subsets and training on these sub-traffic matrices. More specifically, given a topology $G(V, E)$ with $|\Omega|$ SD pairs, each with $K$ paths, the Path-to-Edge matrix has dimensions $(K|\Omega|, |E|)$, and the TE configuration has dimensions $(K|\Omega|, 1)$. Then, while keeping the topology unchanged, we divide the traffic matrix into $S$ sub-traffic matrices, each containing $|\Omega|/S$ SD pairs. Consequently, the dimensions of the Path-to-Edge matrix and TE configuration become $(K|\Omega|/S, |E|)$ and $(K|\Omega|/S, 1)$, respectively. Importantly, we partition only the traffic matrix and not the topology itself. Figure \ref{fig: split_pair} illustrates an example of traffic matrix partitioning in a topology with $3$ nodes and $6$ source-destination pairs. As shown in the figure, the topology remains unchanged, while only the source-destination pairs are partitioned.

\noindent\textbf{Benefits of partitioning.}
Partitioning reduces activation memory and memory consumption from sparse matrix multiplications. Since only a subset of the original traffic matrix is processed, fewer paths are included, leading to lower activation memory. Additionally, smaller sparse matrices require less memory for computation. 

\noindent\textbf{Why partitioning works.}
Equation (\ref{equ: partial e of lagrangian}) shows that updating $\lambda_e$ depends on the difference between edge utilization and the MLU, not on the absolute edge load. As traffic partitioning only changes absolute loads, the update rule remains unaffected, implying that partitioning has a negligible impact on training the dual update operator.
Evaluations in Appendix \ref{sec: visualization of neural network} further confirm that both the distribution of MLP 2’s input and the trained neural network parameters remain highly similar before and after partitioning.

\subsection{Evaluation setup}
\label{sec: evaluation setup}
Our evaluation setup is as follows: The experiments are conducted using Ubuntu 18.04, an Intel Xeon Silver 4110 CPU at 2.10 GHz, and 128 GiB RAM. The programming is carried out using Python 3.8, and the deep learning framework employed is PyTorch, version 2.4.0. The default GPU used is the NVIDIA Tesla P100-PCIE with 16 GiB of VRAM, running CUDA version 11.8. However, due to the high GPU memory demands of TE schemes like HARP, some topologies cause `Out of Memory' (OOM) errors when tested on the Tesla P100-PCIE. Additionally, as analyzed in \S \ref{sec: ml-based TE needs streamlining}, HARP is challenging to parallelize across multiple GPUs. Consequently, for these cases, we have to use the advanced A100-SXM4-80GiB GPU with 80 GiB. To ensure consistency, all time-related results obtained on the A100-SXM4-80GiB are normalized to align with the computation speed of the Tesla P100-PCIE, based on the proportional difference in processing speed between the two GPUs. We use the Tesla P100-PCIE as the primary GPU to showcase that even on a standard GPU, our approach remains both effective and practical.

\subsection{Hyperparameter Search}
\label{sec: Hyperparameter Search}
For DOTE, our hyperparameter search considers a learning rate of (1e-4, 1e-3), the number of hidden layers (2, 3, 4), and a hidden dimension of (128, 256). FIGRET shares the same architectural hyperparameters as DOTE, with an additional robustness-enhancing loss function, for which we set the weight to 0.1.

For TEAL, we explore a learning rate of (1e-5, 1e-4, 5e-4, 1e-3), FlowGNN layers (6, 8), and (5, 10) samples for reward estimation. In addition to hyperparameters, we find that normalizing the input traffic matrix and link capacities improves performance \cite{alqiam2024transferable}. To achieve this, we experiment with multiple renormalization factors. Moreover, TEAL’s primary objective is to maximize the total feasible flow, meaning the total traffic that can be accommodated without exceeding link capacities. As a result, the split ratios for a given source-destination pair do not necessarily sum to 1, reflecting uncollected traffic. However, in MLU optimization, this constraint is required. To align TEAL with MLU optimization, we introduce a minor adjustment to its formulation.

For HARP, we search over a learning rate of (1e-4, 1e-3, 7e-3), the number of RAUs (3–10), GNN layers (1, 2, 4), and Transformer layers (1, 2).

For RedTE, we follow all hyperparameter settings specified in its source code \cite{RedTE}, including random seed and reinforcement learning parameters such as epsilon step, explore decay, and others. For complete details, please refer to \cite{RedTE}.

For SaTE, we follow the hyperparameter settings specified in its source code \cite{SaTE}. In addition, for AlloGNN (which processes the bipartite graph between path-edges and flow-paths), we experiment with hidden sizes (32, 64, 128) and with the number of attention heads (1, 2, 4).

For Geminet, we consider a learning rate of (1e-4, 7e-3), the number of iterative adjustments (3–10), and two MLP components: MLP1 in the Initial Dual Variable Generation Module with (2, 3, 4, 5) layers and MLP2 in the Adjustment Module with (1, 2) layers.
\subsection{Details of DynGeant}
\label{section: details of dyngeant}
\begin{figure}[ht]
    \centering
    \subfigure[Variation in the number of nodes/edges per snapshot (normalized to the maximum number of nodes across snapshots).]{
    \label{fig: dyngeant_node_edge}
    \includegraphics[scale=0.23]{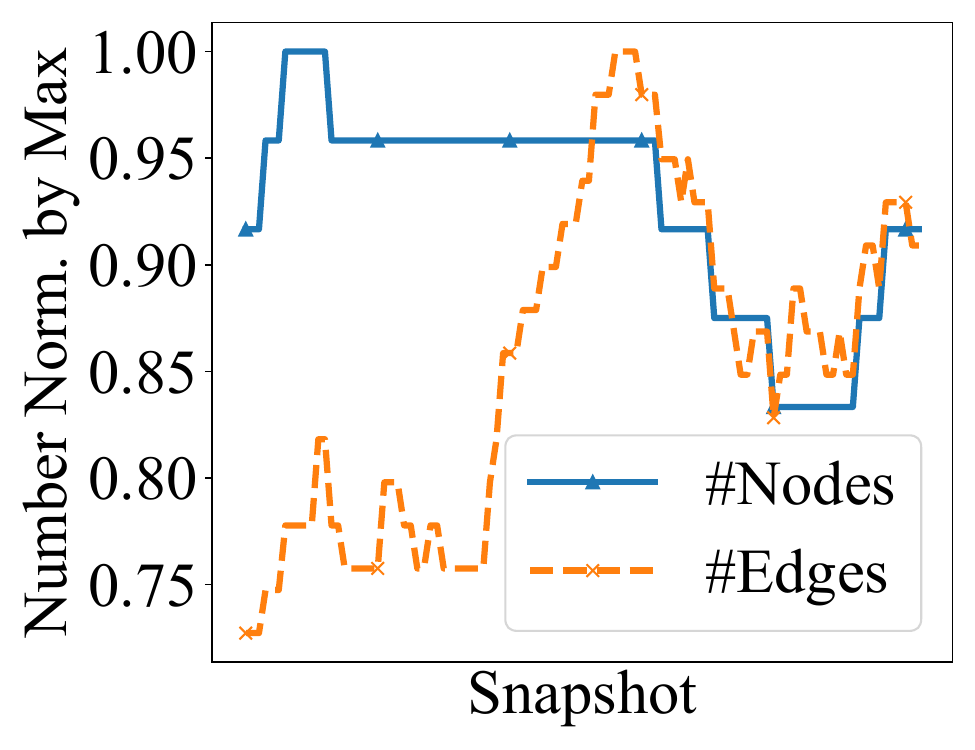}
    }
    \hspace{1mm}
    \subfigure[Changes in Paths across Clusters: Analysis of path additions and removals in clusters at every 5-Cluster interval.]{
    \label{fig: dyngeant_tunnel}
    \includegraphics[scale=0.23]{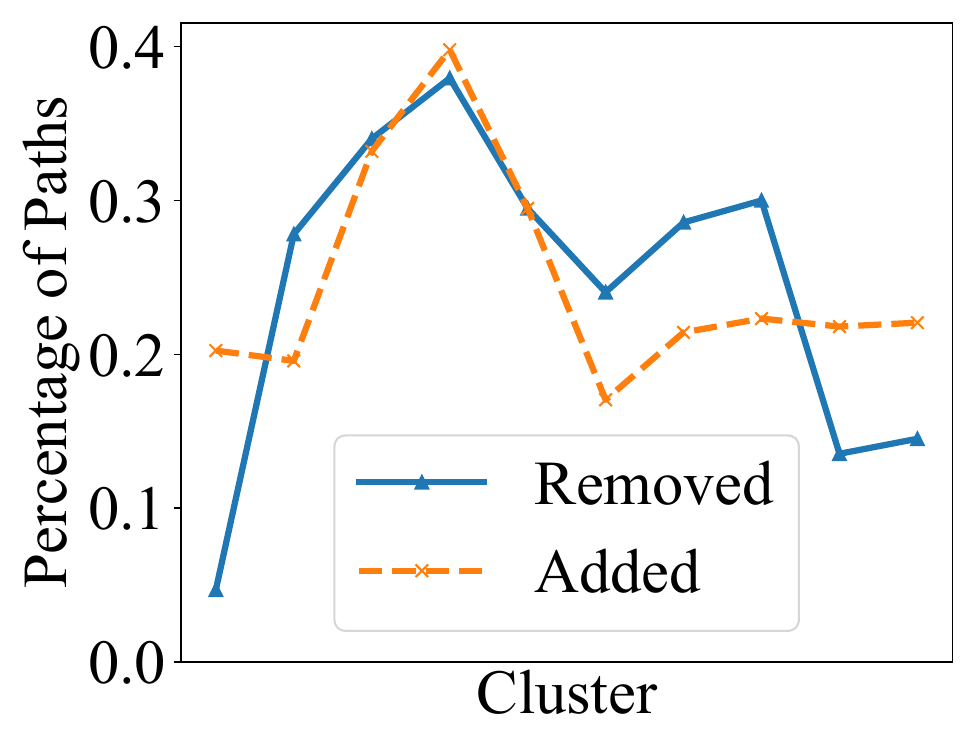}
    }
    \caption{Topology variations across clusters in the DynGeant.}
    \label{fig: variations across clusters}
\end{figure}

\begin{figure}[ht]
    \centering
    \subfigure[Number of unique capacity values per link across snapshots of the last cluster.]{
    \label{fig: dyngeant_capacity}
    \includegraphics[scale=0.23]{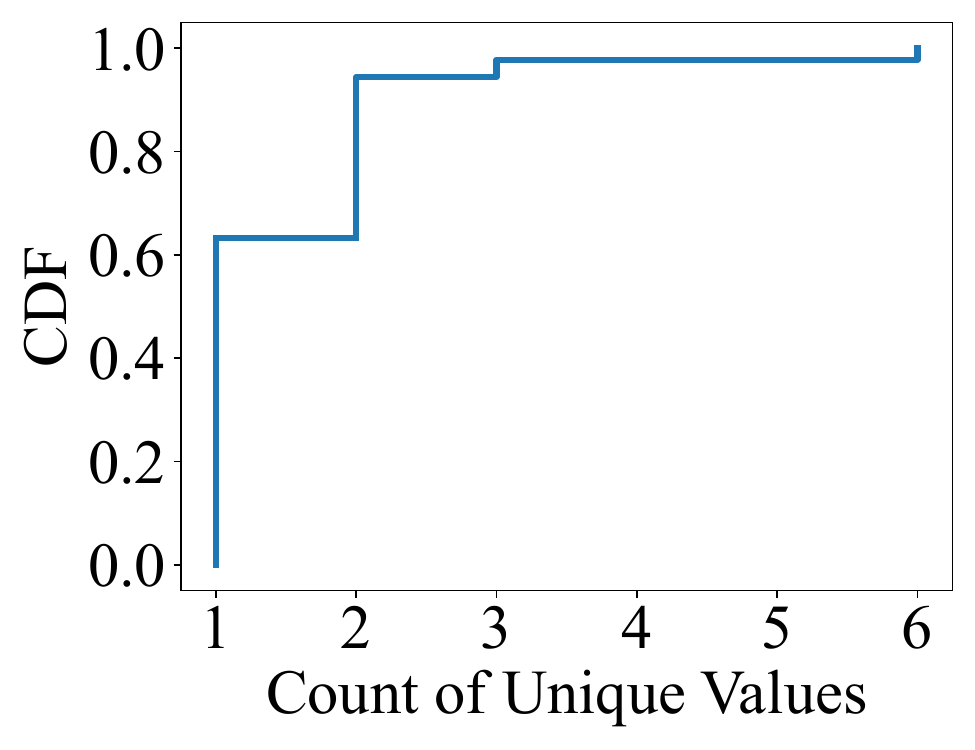}
    }
    \hspace{1mm}
    \subfigure[Average capacity across snapshots of the last cluster.]{
    \label{fig: dyngeant_avg_capacity}
    \includegraphics[scale=0.23]{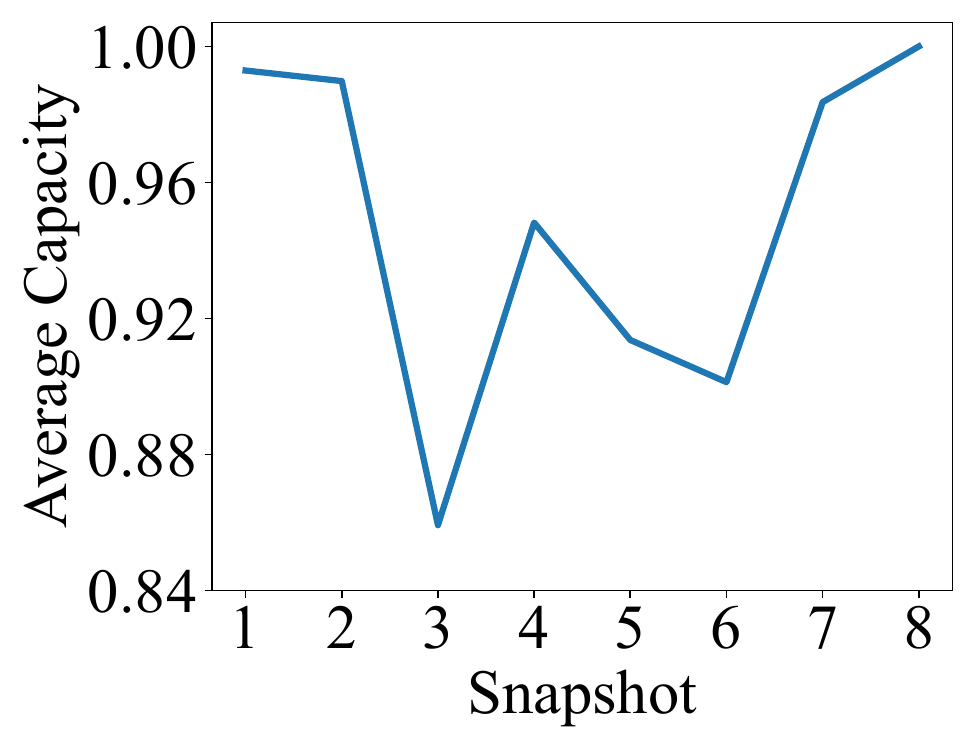}
    }
    \caption{Topology variations across snapshots within the last cluster in the DynGeant.}
    \label{fig: variations within clusters}
\end{figure}

In this section, we present the topological variations of DynGeant. The topological variations across clusters and within clusters are depicted in Figure \ref{fig: variations across clusters} and Figure \ref{fig: variations within clusters}, respectively.
\label{sec: details of dyngeatn}

\subsection{Details of memory overhead in RedTE}
\label{sec: redte_memory}
We report the memory usage of all TE schemes except RedTE in Table~\ref{tab: memory_usage}. In our experiments, RedTE's memory usage exceeds 80~GiB on all topologies except GEANT. To illustrate the root cause, we analyze RedTE's model memory usage theoretically based on its open-source implementation \cite{RedTE}. In RedTE, the critic network takes two high-dimensional inputs: (i) the splitting ratios for all source--destination pairs and (ii) the aggregated local environment features of all nodes. On the KDL topology with 4 paths per source--destination pair, the input dimensionalities are 2{,}271{,}048 and 569{,}552, respectively. According to the open-source implementation, each critic contains 219 million parameters. Using float32, a single critic requires $219\times 10^6 \times 4 \approx 0.88$~GiB to store parameters alone. RedTE maintains a critic for each agent during training, so storing all critic models requires approximately 613~GiB of memory (even before accounting for optimizer states, gradients, and activation buffers), highlighting RedTE's poor scalability to large topologies.
\section{Additional Results}
\label{sec: additional results}

\subsection{Ablation Study}
\label{sec: Ablation Study}
\begin{figure}[!ht]
    \centering
    \includegraphics[scale=0.23]{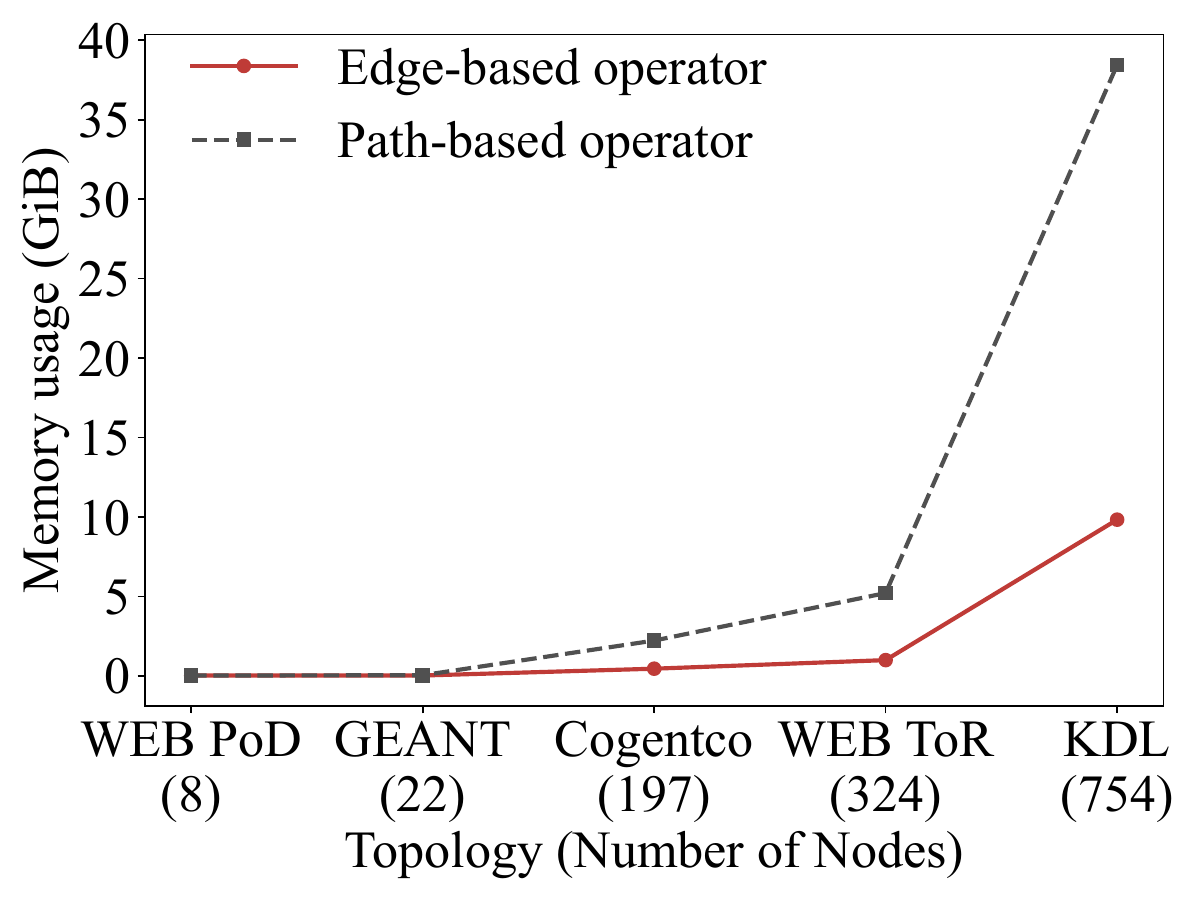}
    \caption{Comparison of memory usage between Path-based operator and Edge-based operator.}
    \label{fig: memory_usage_path_vs_edge}
\end{figure}

We conduct ablation studies to examine the effectiveness of \S \ref{sec: learning the iterative process} and \S \ref{sec: harness duality in TE}. Specifically, we compare the Path-based update operator and the Edge-based update operator in terms of memory consumption and TE performance, with the results presented in Figure \ref{fig: memory_usage_path_vs_edge} and Table \ref{tab: performance_path_edge}. The observations are as follows: (i) both operators achieve the same TE performance; (ii) regardless of which operator is used for iteration, the memory consumption remains consistently lower than that of previous approaches, with the edge-based update operator consuming less memory.

\begin{table}[!ht]
\scalebox{0.74}{
\begin{tabular}{c|ccccc}
\toprule
Topology name     & WEB PoD & GEANT & Cogentco & WEB ToR & Kdl  \\ \hline
Performance Ratio & 0.99    & 1  & 1     & 0.99    & 1 \\ \bottomrule
\end{tabular}
}
\caption{Path-based operator performance, where the ratio is its MLU divided by the Edge-based operator’s.}
\label{tab: performance_path_edge}
\end{table}

\begin{figure*}
    \begin{minipage}{0.32\textwidth}
    \includegraphics[scale=0.3]{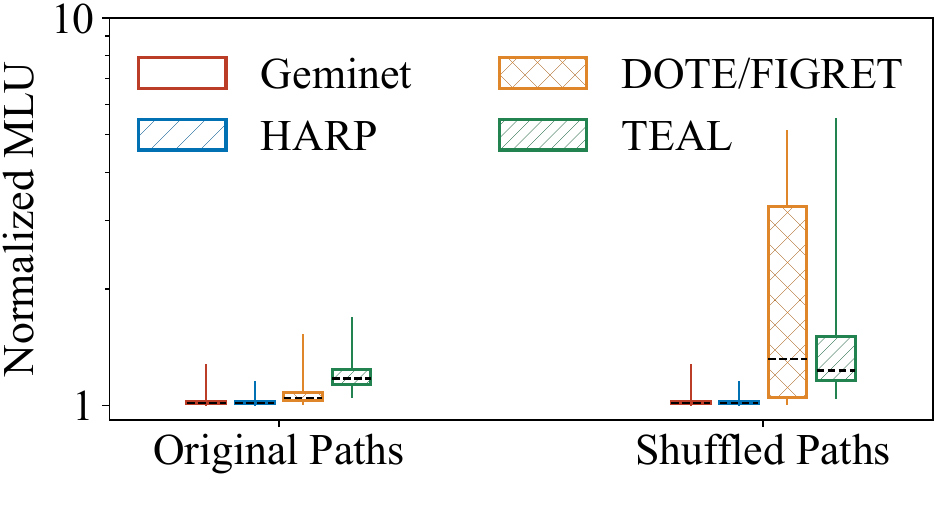}
    \caption{Comparing ML-based TE schemes for PoD-level Meta WEB.}
    \label{fig: one_cluster_pod_b}
    \end{minipage}
    \hspace{2mm}
    \begin{minipage}{0.32\textwidth}
    \includegraphics[scale=0.3]{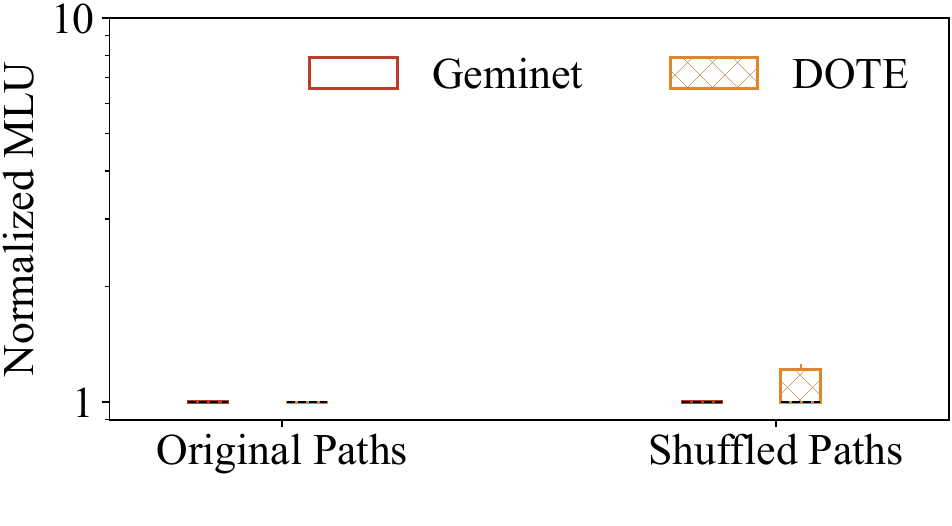}
    \caption{Comparing ML-based TE schemes for KDL topology.}
    \label{fig: one_cluster_kdl}
    \end{minipage}
    \hspace{2mm}
    \begin{minipage}{0.32\textwidth}
    \includegraphics[scale=0.3]{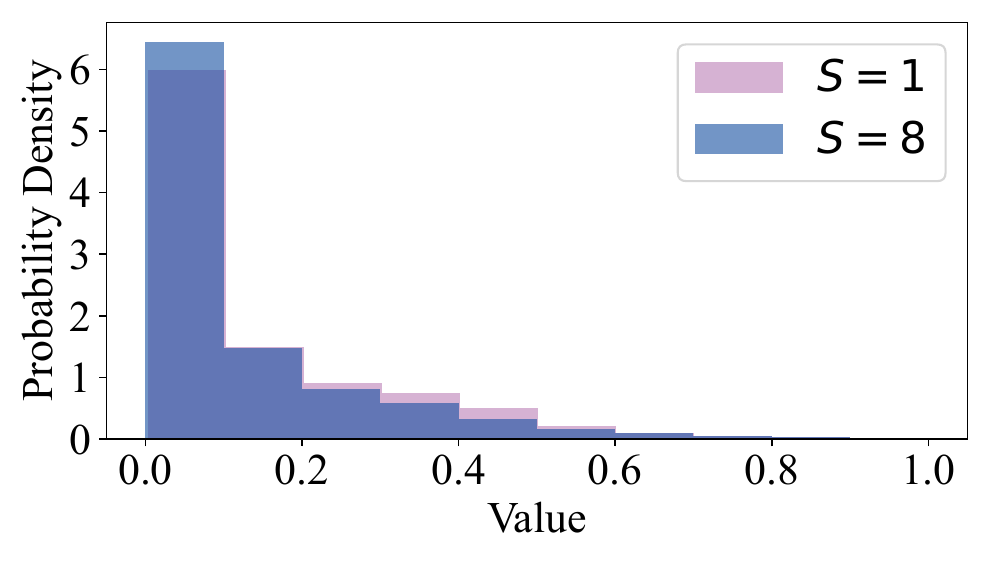}
    \caption{The distribution of edge utilization divided by MLU.}
    \label{fig: input distribution}
    \end{minipage}
\end{figure*}

\begin{figure*}[!ht]
    \centering
    \subfigure[$S=1$: Input layer.]{
    \label{fig: input layer}
    \includegraphics[scale=0.18]{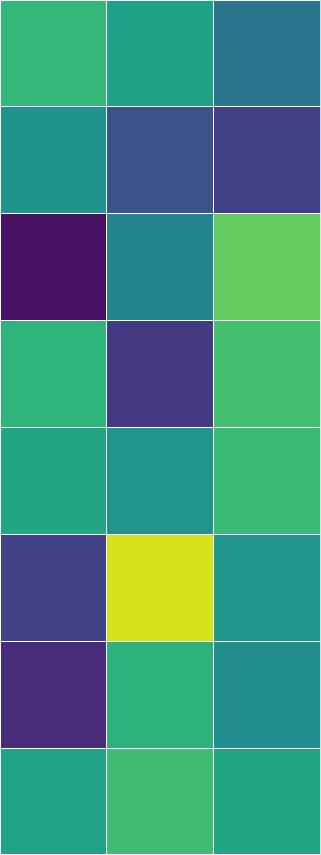}
    }
    \subfigure[$S=1$: Hidden layer.]{
    \label{fig: hidden layer}
    \includegraphics[scale=0.18]{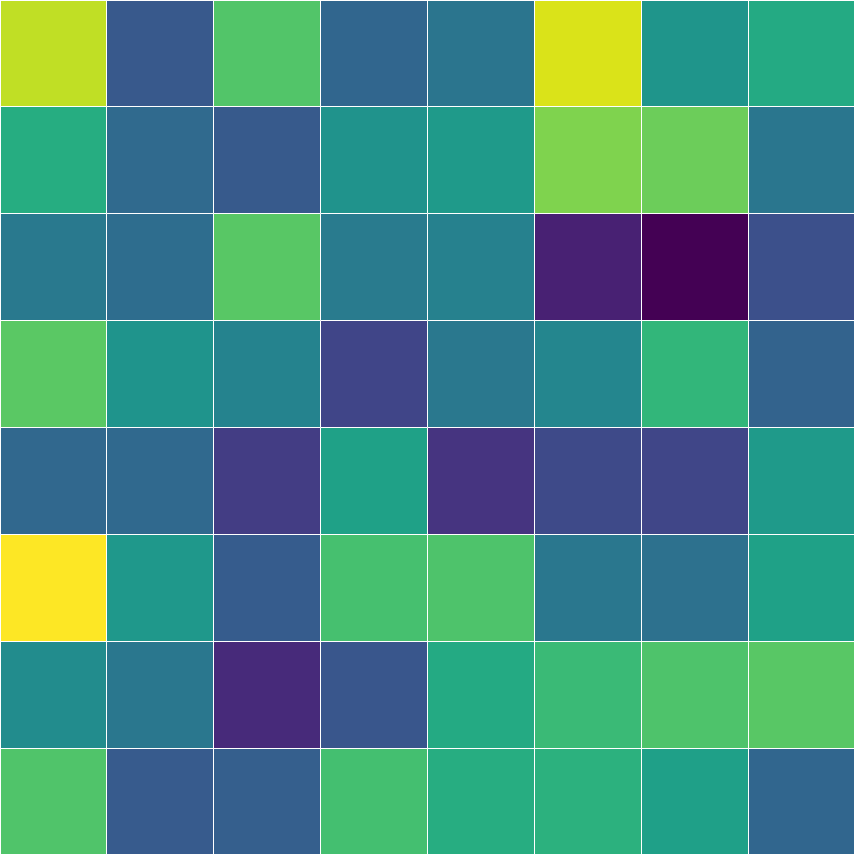}
    }
    \subfigure[$S=1$: Output layer]{
    \label{fig: output layer}
    \includegraphics[scale=0.18]{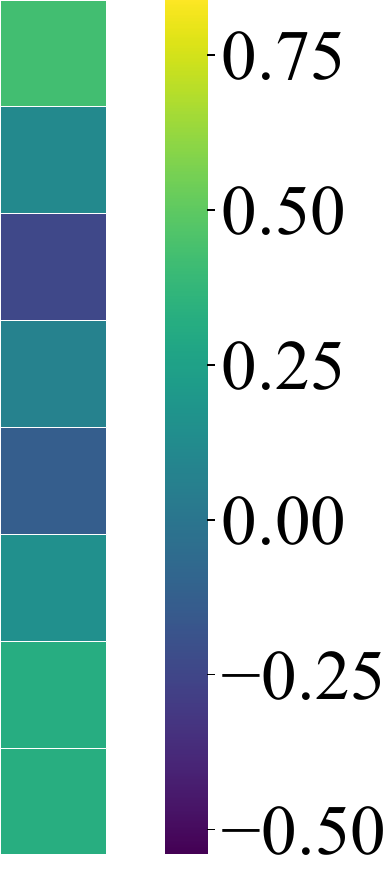}
    }
    \hspace{8mm}
    \subfigure[$S=8$: Input layer.]{
    \label{fig: split input layer}
    \includegraphics[scale=0.18]{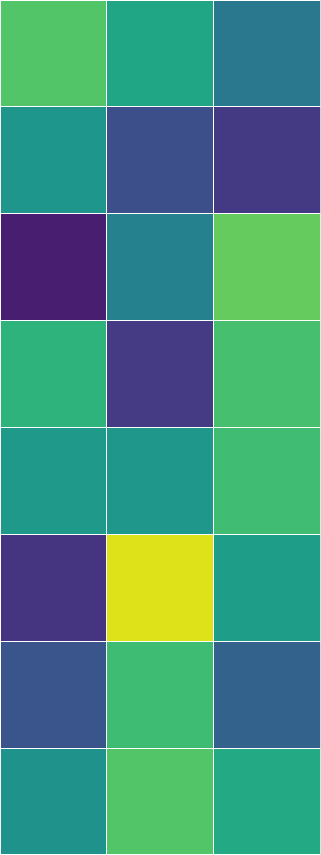}
    }
    \subfigure[$S=8$: Hidden layer.]{
    \label{fig: split hidden layer}
    \includegraphics[scale=0.18]{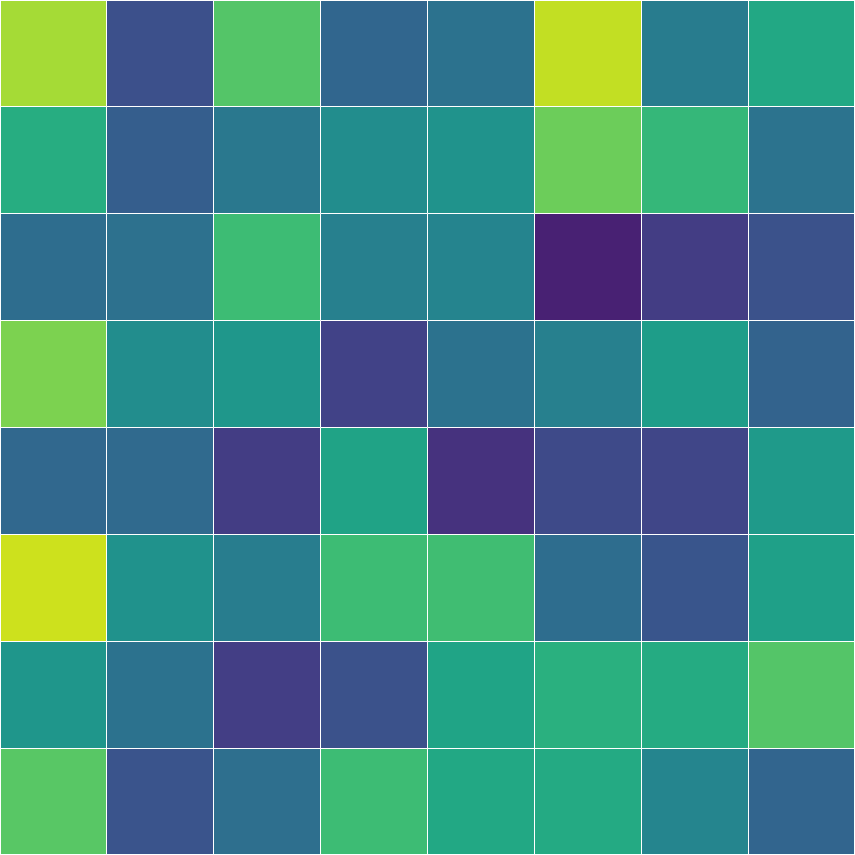}
    }
    \subfigure[$S=8$: Output layer]{
    \label{fig: split output layer}
    \includegraphics[scale=0.18]{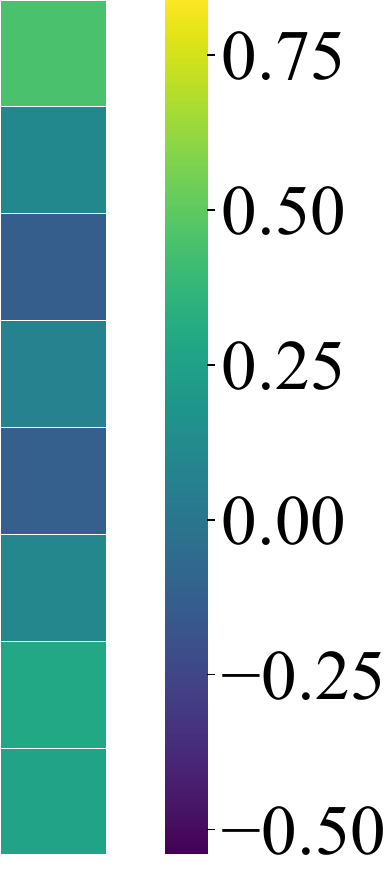}
    }
    \caption{Comparison of the neural network parameters between the non-partitioned baseline $S=1$ and the model trained with partitioning into 8 groups $S=8$ (using only one group for training) on KDL. For better visualization, the input layer weight matrix has been transposed. The trained neural network exhibits highly similar parameters.}
    \label{fig: visualization of neural network}
\end{figure*}

\subsection{Additional performance results}
\label{sec: additional performance results}
This section presents additional performance results evaluated on the one-cluster topologies. This includes results on the PoD-level Meta WEB topology (Figure \ref{fig: one_cluster_pod_b}) and on the KDL topology (Figure \ref{fig: one_cluster_kdl}). In Figure \ref{fig: one_cluster_kdl}, considering the low throughput of HARP and TEAL on KDL, which results in long training times, our results only include Geminet and DOTE. However, this does not affect our conclusion: on the original tunnels, Geminet performs on par with any ML-based TE scheme and can handle tunnel shuffle.

\subsection{Training on sub traffic matrix}
\label{sec: training on sub-traffic}
\begin{figure}[!ht]
    \centering
    \includegraphics[width=0.6\linewidth]{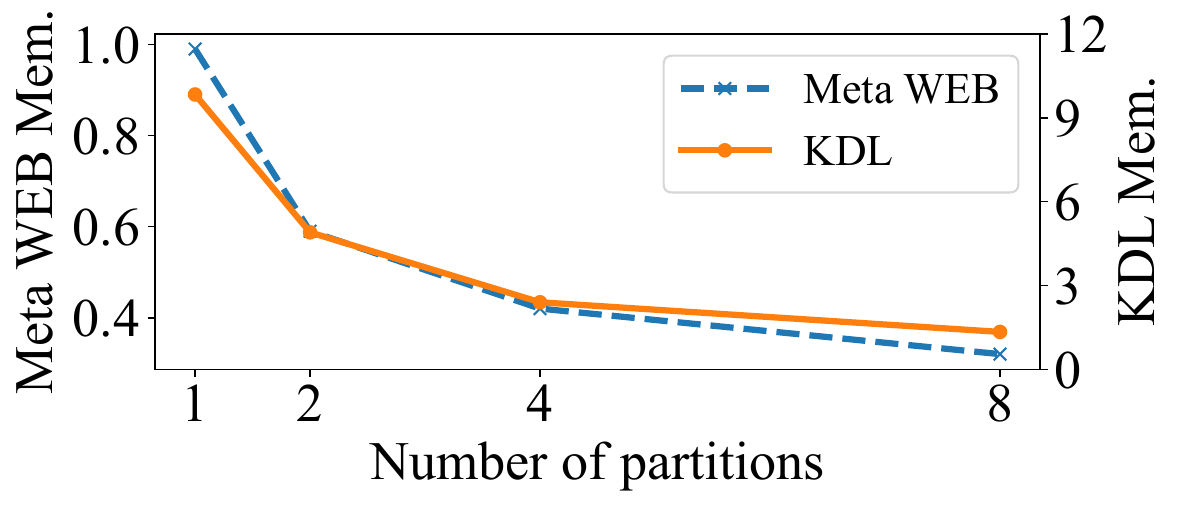}
    \caption{GPU memory usage (GiB) with partitions ($S$).}
    \label{fig: split_memory}
\end{figure}

\noindent\textbf{Memory usage.}  Figure~\ref{fig: split_memory} reports GPU memory consumption during training. As the number of partitions increases, peak memory usage decreases substantially. For example, on the KDL dataset, using two partitions reduces memory consumption to $49.8\%$ of the non-partitioned baseline, while four partitions further reduce it to $24.4\%$.

\begin{figure}[!ht]
    \centering
    \includegraphics[scale=0.28]{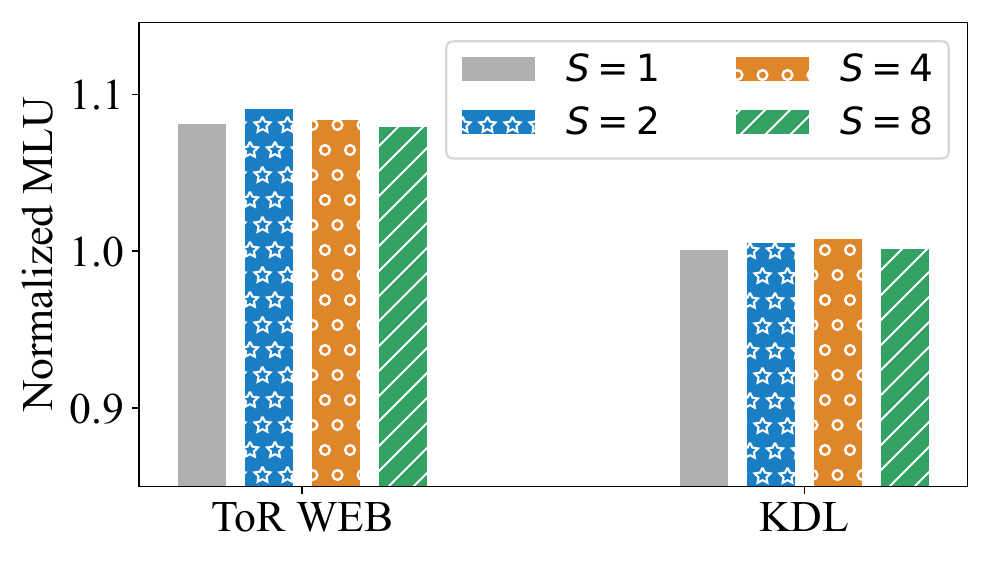}
    \caption{Changes in average normalized MLU with the number of partitions ($S$) when training with just one of the partitioned groups.}
    \label{fig: training on one group}
\end{figure}

\noindent\textbf{Performance of training on one group.} Figure~\ref{fig: training on one group} shows the performance of this. Even when training with a single partitioned group, Geminet maintains robust performance with only marginal degradation: $0.7\%$, $0.1\%$, and $0.09\%$ drops on ToR WEB for $S=2,4,8$, respectively; and $0.4\%$, $0.7\%$, and $0.05\%$ drops on KDL.

\begin{figure}[!ht]
    \centering
    \includegraphics[scale=0.28]{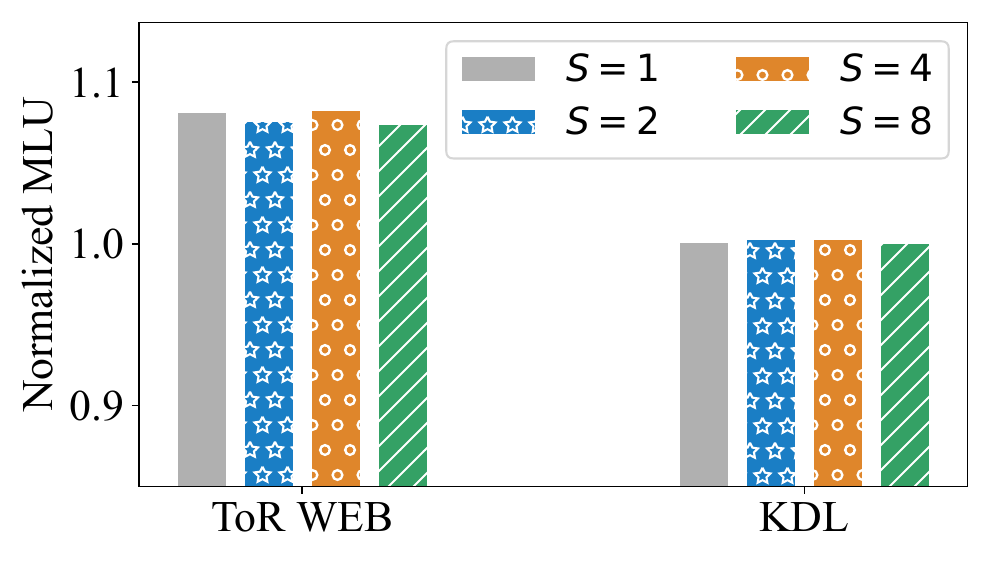}
    \caption{Changes in average normalized MLU with the number of partitions ($S$) when training with all groups.}
    \label{fig: training on all groups}
\end{figure}

\noindent\textbf{Performance of training on all groups.} Figure \ref{fig: training on all groups} summarizes the performance results when training with all partitioned groups. From the results, it can be observed that training with all partitioned groups performs as well as the non-partitioned baseline. On the WEB ToR topology, the performance drops compared to the non-partitioned baseline for $S=2$, $S=4$, and $S=8$ are $-0.7\%$, $-0.1\%$, and $-0.8\%$, respectively. For the KDL topology, the corresponding values are $0.1\%$, $0.1\%$, and $-0.1\%$. Negative values indicate an improvement rather than a drop in performance.

Regarding why the performance improves, our initial investigation suggests that when the data is partitioned into $s$ groups, it adds $s-1$ additional groups of training data compared to the non-partitioned scenario.
As discussed in Appendix \ref{sec: training on sub-pairs}, the edge relative load (edge utilization / Max Link Utilization) is more critical for adjustments. The additional $s-1$ groups of training data may expose the neural network to more diverse edge relative load scenarios, thereby enhancing the model's ability to make adjustments.

\subsection{Interpreting of training on sub-tms}
\label{sec: visualization of neural network}
In this section, we visualize the distribution of the input values and the parameters of the neural network models.
Figure \ref{fig: input distribution} shows the distribution of relative link utilization (edge utilization divided by MLU) before and after partitioning. It can be observed that partitioning has little impact on the distribution of relative link utilization.
As analyzed in \S \ref{sec: training on sub-pairs}, Geminet relies on relative link utilization rather than absolute edge loads for training. Since partitioning does not alter the relative utilization distribution, it does not affect the final training performance.

To further validate this, we visualize the trained neural network parameters. Figure \ref{fig: visualization of neural network} compares the model of the non-partitioned baseline with the model trained on one group after partitioning into 8 groups for the KDL topology. The neural networks consist of three layers: an input layer, a hidden layer, and an output layer. The weight matrices of these three layers are visualized. 
From the results, it can be observed that the parameters of the trained neural networks are relatively similar. The cosine similarity is $97\%$.
\subsection{Additional Traffic Uncertainty Results}
\label{sec: additional traffic uncertainty results}
\begin{figure}
    \centering
    \includegraphics[width=0.6\linewidth]{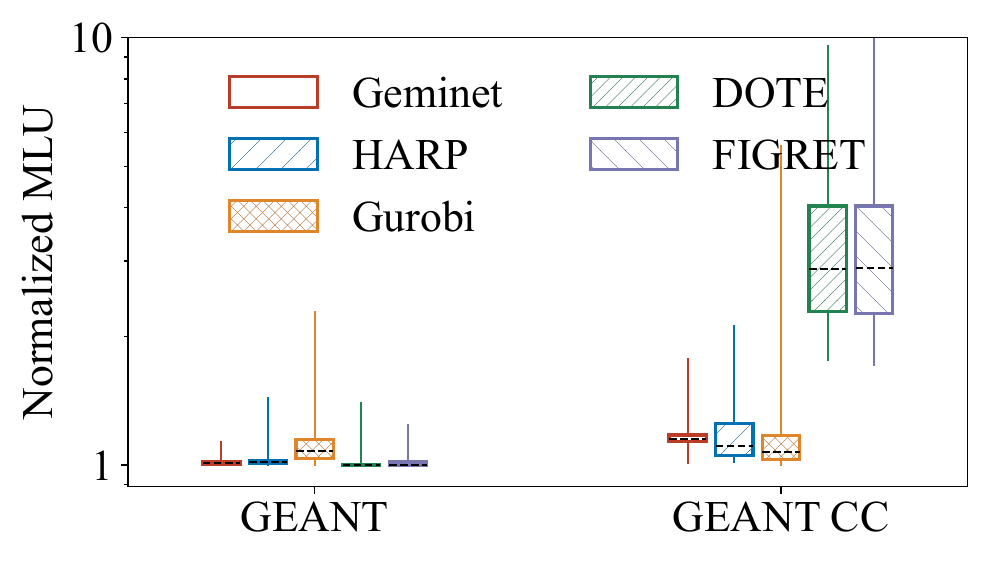}
    \caption{Comparison of schemes on GEANT under traffic uncertainty, where CC represents changing edge capacity.}
    \label{fig: geant traffic uncertainty}
\end{figure}

We present additional results on handling traffic uncertainty in the WAN topology GEANT. Figure~\ref{fig: geant traffic uncertainty} compares different schemes under traffic uncertainty on GEANT, where \textsc{CC} denotes scenarios with changing edge capacity.

\subsection{Fewer parameters, less training data}
\label{sec: fewer parameters less training data}
\begin{figure}[!ht]
\begin{minipage}{0.23\textwidth}
\centering
    \includegraphics[scale=0.19]{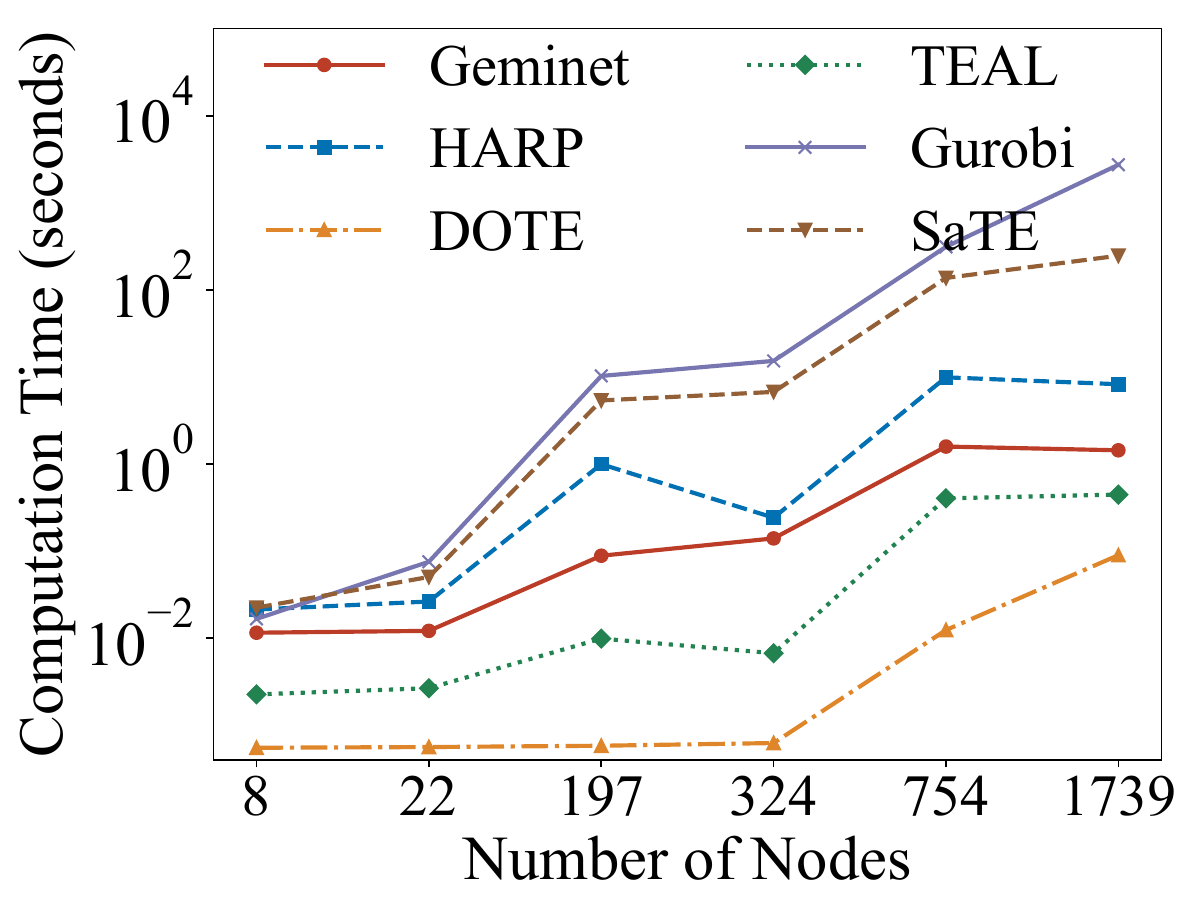}
    \caption{Comparison of computation times across different topology sizes.}
    \label{fig: comparison of computation times}
\end{minipage}
\hspace{1mm}
\begin{minipage}{0.23\textwidth}
    \centering
    \includegraphics[scale=0.19]{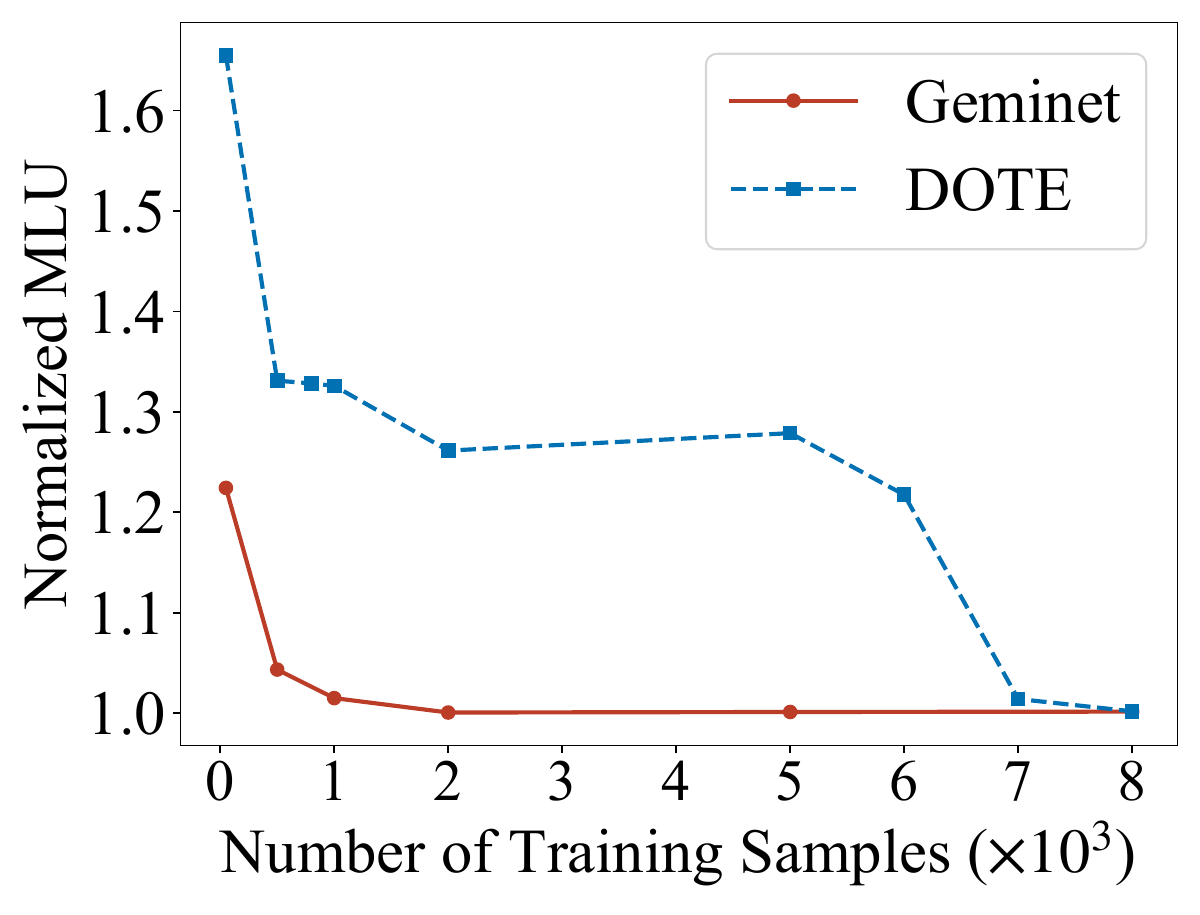}
    \caption{Relationship between model quality and data quantity on GEANT.}
    \label{fig: few_training}
\end{minipage}
\end{figure}

In this section, we compare Geminet, a TE scheme with fewer parameters, to DOTE, which has significantly more parameters, under reduced training data conditions to highlight the advantage of models with fewer parameters requiring less data to train effectively. The results, summarized in Figure~\ref{fig: few_training}, illustrate the relationship between model quality and training data quantity for Geminet and DOTE. Geminet requires less training data than DOTE due to its smaller number of parameters—only 20\% of the data is sufficient.

\subsection{Additional link failure results}
\label{sec: additional link failure results}
\begin{figure}[!ht]
    \centering
    \includegraphics[scale=0.32]{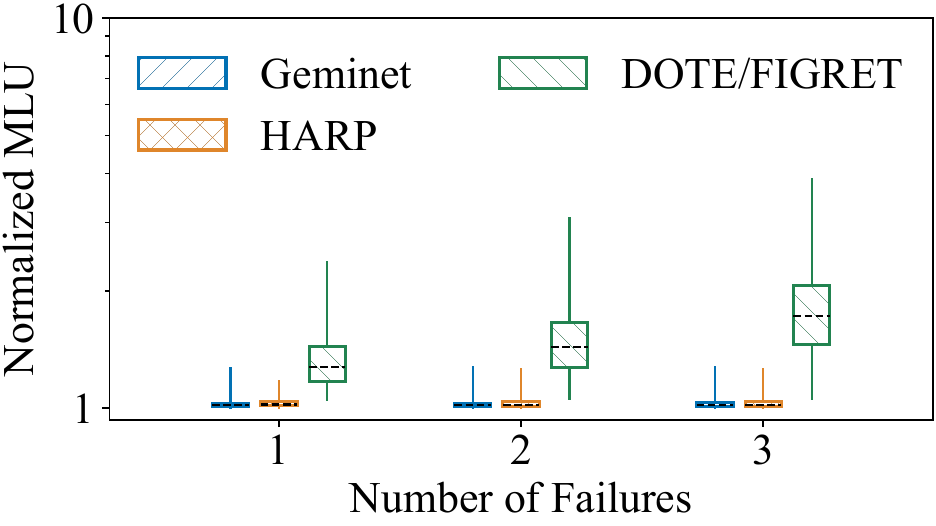}
    \caption{Coping with different numbers of random link failures on PoD-level Meta WEB.}
    \label{fig: pod b failure}
\end{figure}

\begin{figure}[!ht]
    \centering
    \includegraphics[scale=0.32]{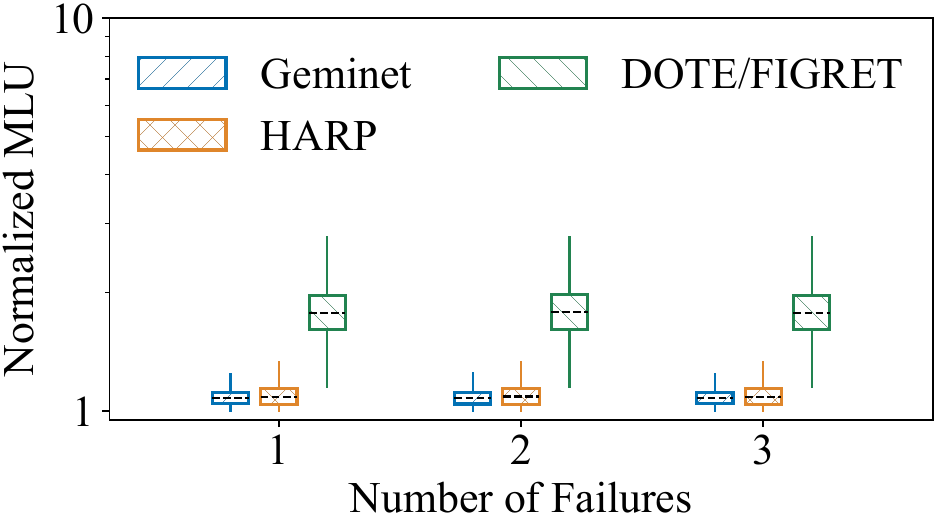}
    \caption{Coping with different numbers of random link failures on ToR-level Meta WEB.}
    \label{fig: tor b failure}
\end{figure}

In this section, we present the results from evaluating link failures on PoD-level Meta WEB and ToR-level Meta WEB, as shown in Figure \ref{fig: pod b failure} and Figure \ref{fig: tor b failure}. 
From the results, it can be observed that Geminet performs on par with HARP across WAN topologies, data center PoD-level, and ToR-level topologies. Additionally, Geminet and HARP consistently outperform DOTE in addressing link failures.

\subsection{Inference Memory Footprint}
\label{sec: inference_memory}
Table~\ref{tab: inference_memory} summarizes Geminet's peak GPU memory footprint during inference.

\begin{table}[t]
\centering
\scalebox{0.8}{
\begin{tabular}{ccccccc}
\hline
Topology & 8 & 22 & 197 & 324 & 754 & 1739 \\
\hline
Memory & 8.9e-3 & 9.1e-3 & 1.6e-1 & 9.8e-2 & 4.3 & 4 \\
\hline
\end{tabular}
}
\caption{Peak inference memory (GB).}
\label{tab: inference_memory}
\end{table}

\subsection{Cross-topology generalization results}
\label{sec: reverse cross-topology}

\begin{figure}[!ht]
    \centering
    \includegraphics[width=0.5\linewidth]{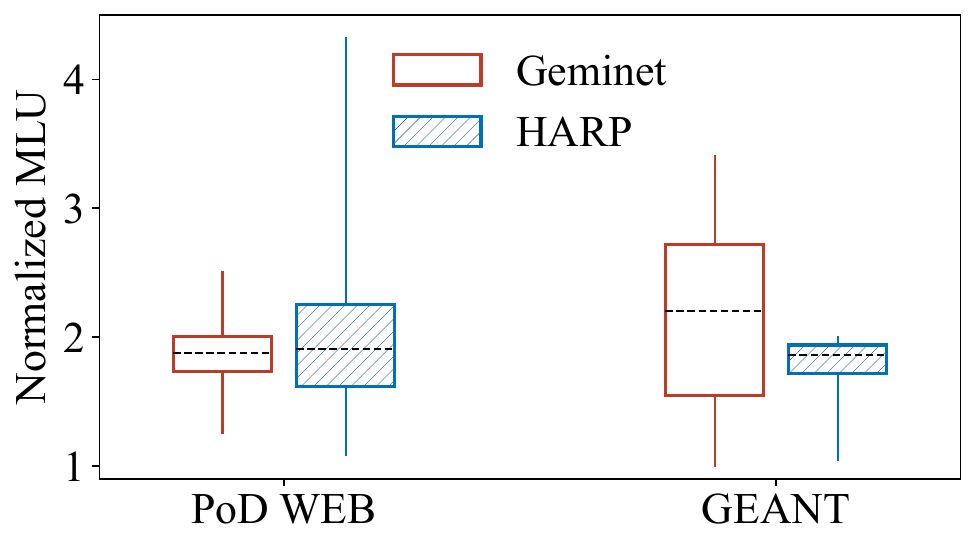}
    \caption{Cross-topology generalization results with models trained on Cogentco and tested on different topologies.}
    \label{fig: reverse_pod_cogentco_comparison}
\end{figure}

To complement \S\ref{sec: handling drastically different topology}, we report an additional transfer setting where the source and target topologies are swapped. Specifically, we train Geminet and HARP on Cogentco and directly test the trained models on PoD WEB and GEANT without any fine-tuning. The results are summarized in Figure~\ref{fig: reverse_pod_cogentco_comparison}.

\subsection{Controlled Input Sweep of MLP 2}
\label{sec: edge_mlp_probe}

\begin{figure}[!ht]
    \centering

    \subfigure[Meta WEB PoD-level.]{
    \label{fig: meta web pod-level update}
    \includegraphics[scale=0.16]{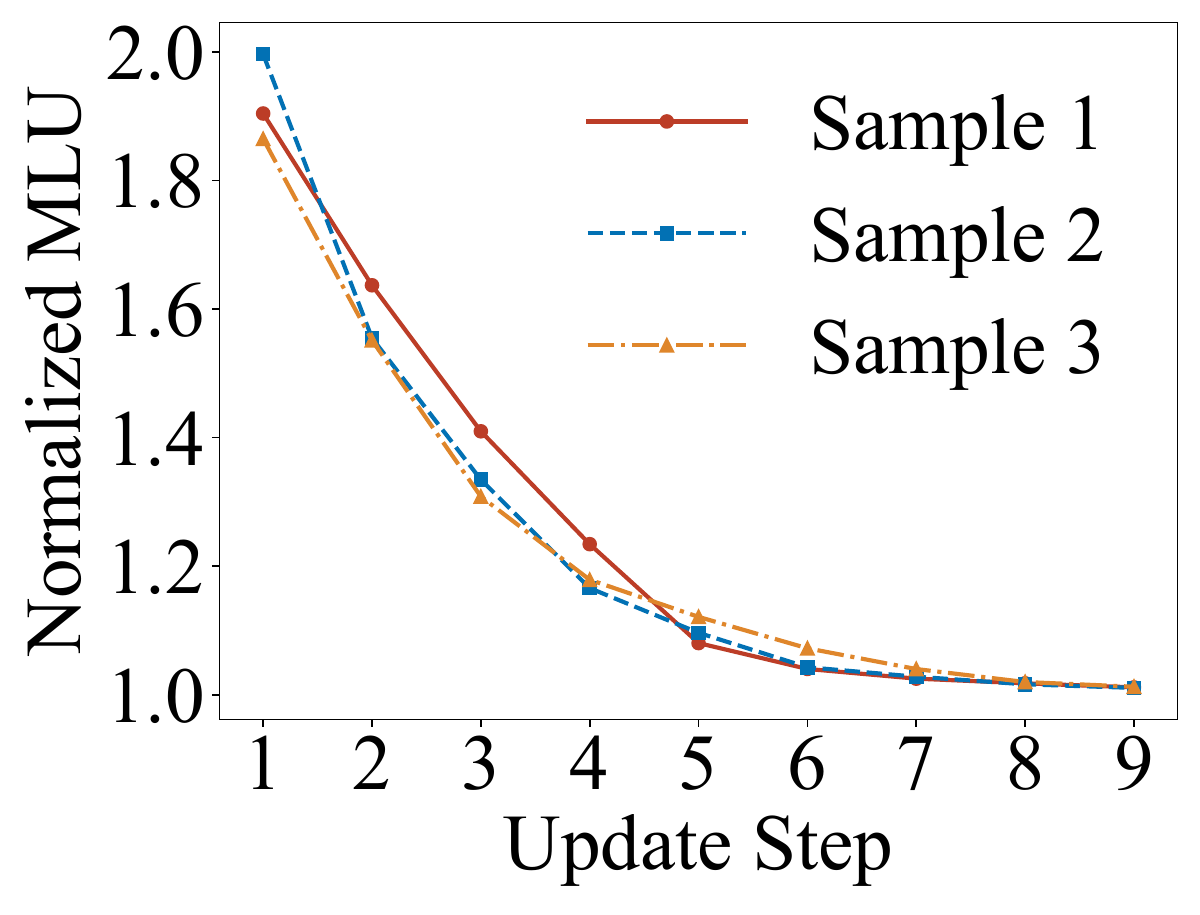}
    }
    \hspace{1mm}
    \centering
    \subfigure[GEANT.]{
    \label{fig: GEANT update}
    \includegraphics[scale=0.16]{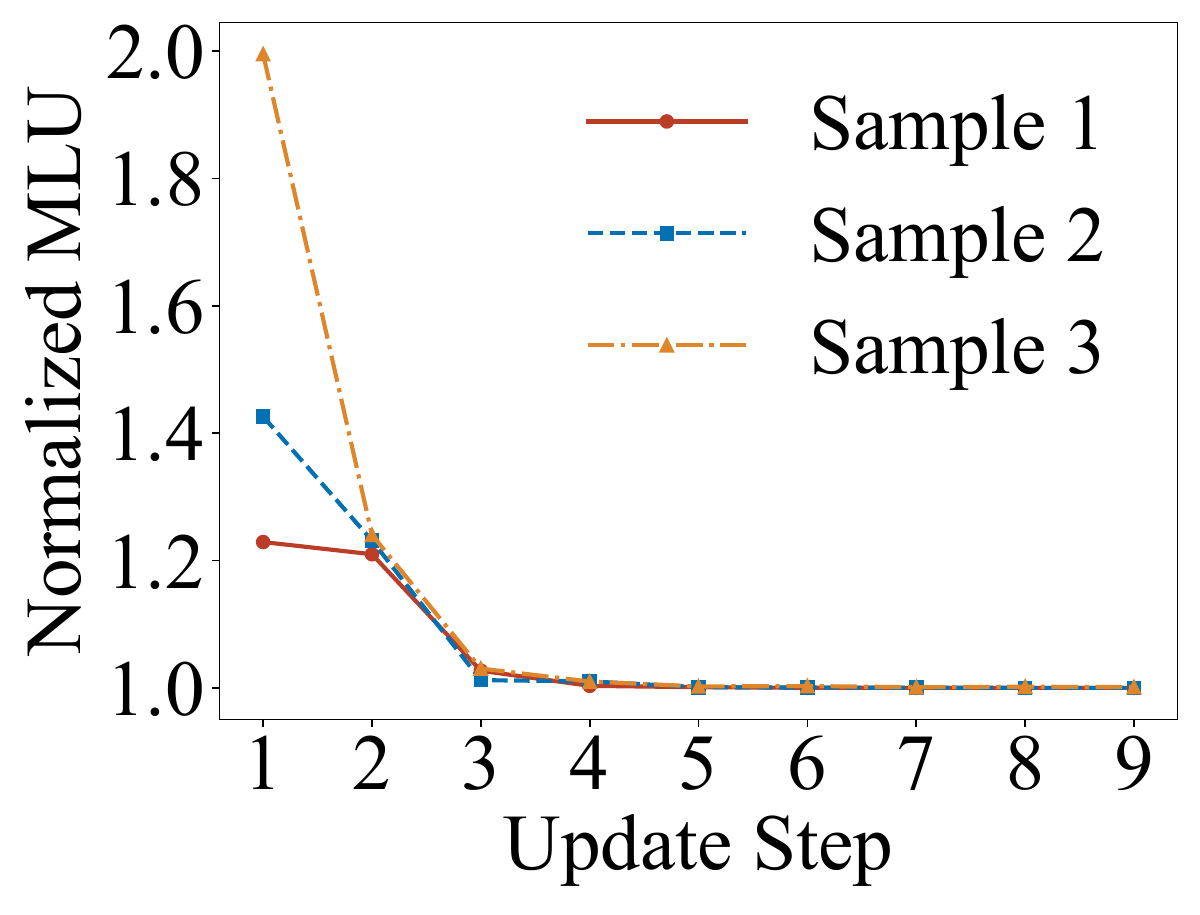}
    }
    \caption{Visualization of the update process.}
    \label{fig: update curve}
\end{figure}

\begin{figure}[!ht]
    \centering

    \subfigure[Meta WEB PoD-level.]{
    \label{fig: meta web pod-level update}
    \includegraphics[scale=0.16]{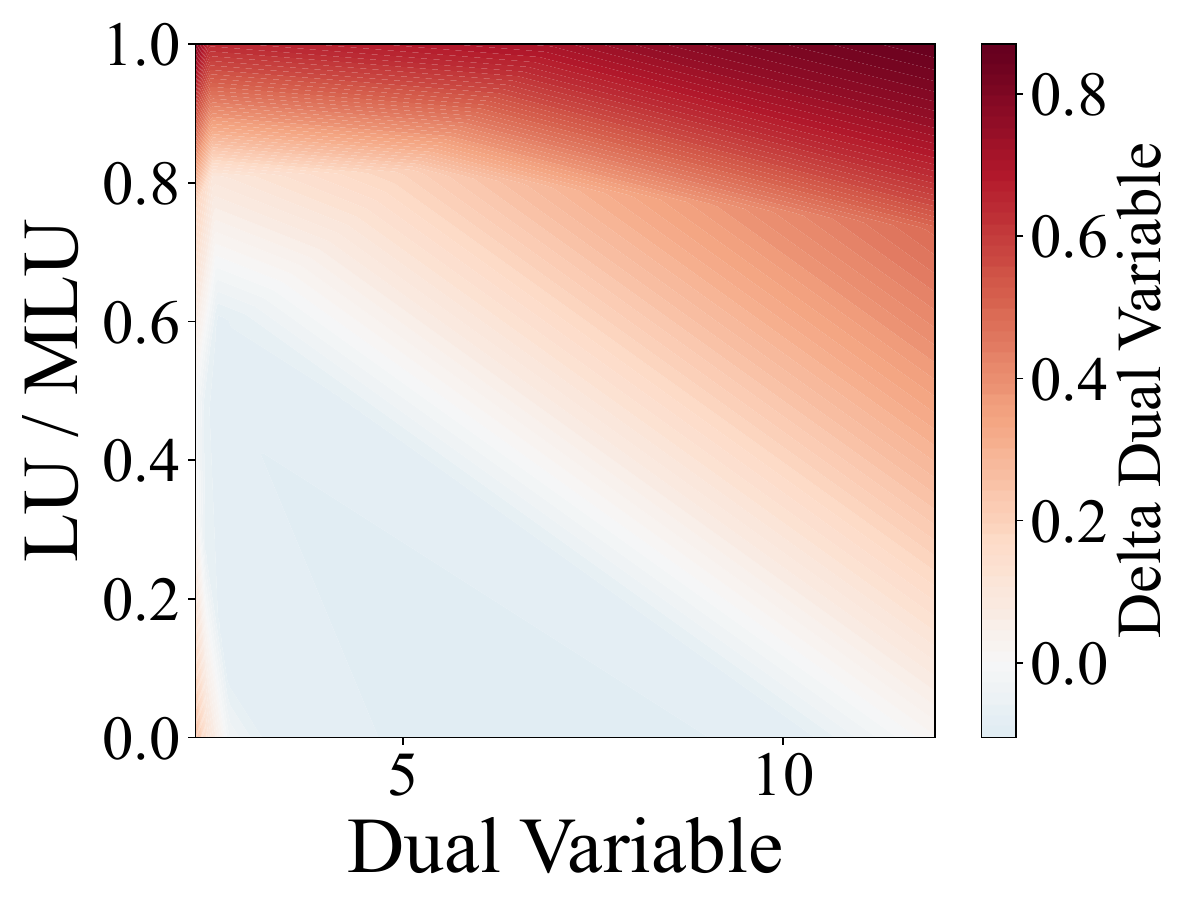}
    }
    \hspace{1mm}
    \centering
    \subfigure[GEANT.]{
    \label{fig: GEANT update}
    \includegraphics[scale=0.16]{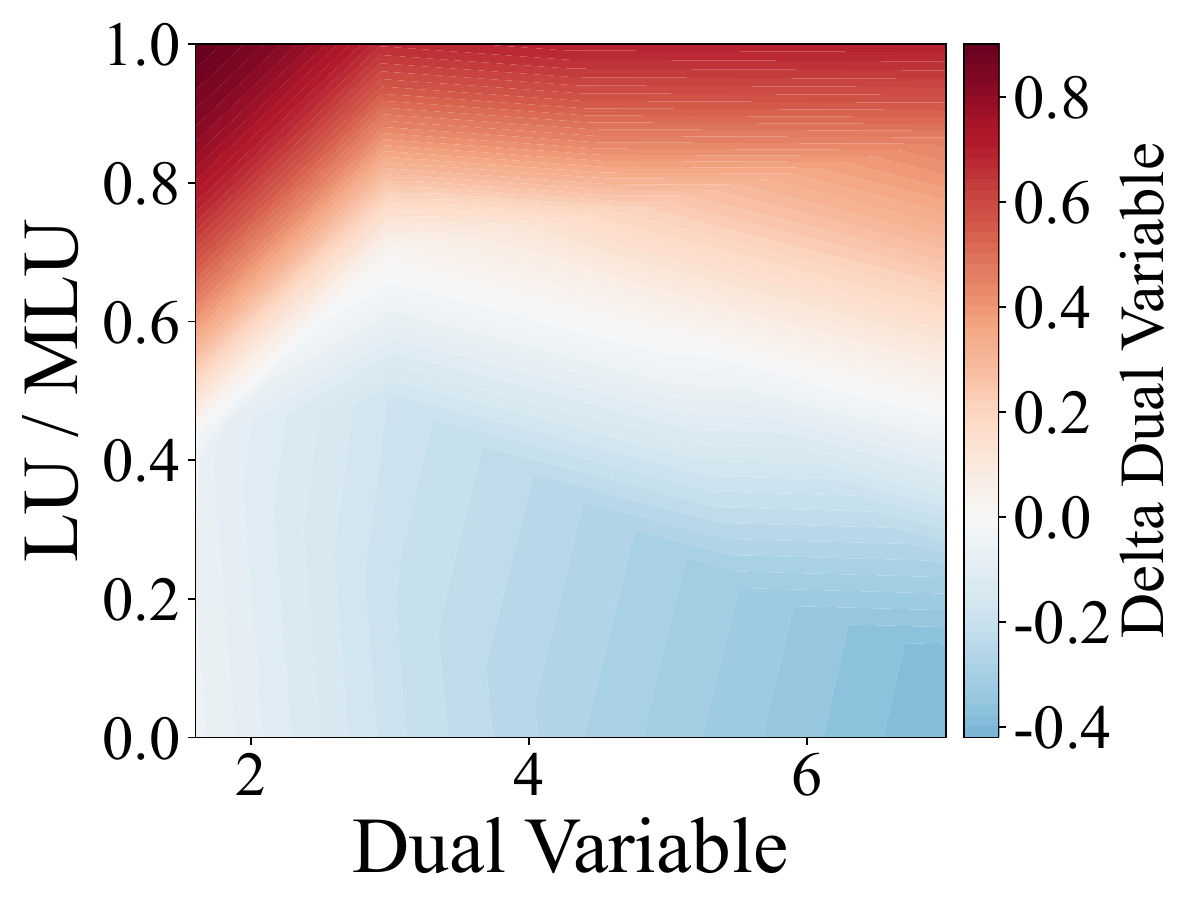}
    }
    \caption{Controlled input sweep of MLP 2 with fixed capacity. We jointly vary $\mathrm{LU}/\mathrm{MLU}$ and $\lambda$ (ranges calibrated from real inference trajectories) and visualize the update behavior.}
    \label{fig: edge_mlp_probe}
\end{figure}

\end{document}